\documentclass[a4paper,12pt]{article}
\pdfoutput=1
\usepackage{graphicx}
\usepackage[margin=1in]{geometry}
\usepackage{amsmath}

\begin{document}
\makeatletter
\@addtoreset{equation}{section}
\makeatother
\renewcommand{\theequation}{\thesection.\arabic{equation}}
\vspace{1.8truecm}

{\LARGE{ \centerline{\bf Structure constants of heavy operators}\par
\centerline{\bf in ABJM/ABJ Theory}  }}  

\vskip.5cm 

\thispagestyle{empty} 
\centerline{ {\large\bf Gaoli Chen$^{b,}$\footnote{{\tt galic.chen@gmail.com}}, 
Robert de Mello Koch$^{a,b,}$\footnote{{\tt robert@neo.phys.wits.ac.za}}, 
Minkyoo Kim$^{b,}$\footnote{{\tt minkyoo.kim@wits.ac.za}} }}\par
\vspace{.2cm}
\centerline{{\large\bf and
Hendrik J.R. Van Zyl${}^{b,}$\footnote{ {\tt hjrvanzyl@gmail.com}} }}

\vspace{.4cm}
\centerline{{\it ${}^{a}$ School of Physics and Telecommunication Engineering},}
\centerline{{ \it South China Normal University, Guangzhou 510006, China}}

\vspace{.4cm}
\centerline{{\it ${}^{b}$ National Institute for Theoretical Physics,}}
\centerline{{\it School of Physics and Mandelstam Institute for Theoretical Physics,}}
\centerline{{\it University of the Witwatersrand, Wits, 2050, } }
\centerline{{\it South Africa } }

\vspace{1truecm}

\thispagestyle{empty}

\centerline{\bf ABSTRACT}

\vskip.2cm 
Efficient and powerful approaches to the computation of correlation functions involving determinant, sub-determinant
and permanent operators, as well as traces, have recently been developed in the setting of ${\cal N}=4$ super Yang-Mills theory.
In this article we show that they can be extended to ABJM and ABJ theory. 
After making use of a novel identity which follows from character orthogonality, an integral representation of certain
projection operators used to define Schur polynomials is given.
This integral representation provides an effective description of the correlation functions of interest.
The resulting effective descriptions have ${1\over N}$ as the loop counting parameter, strongly suggesting their
relevance for holography.

\setcounter{page}{0}
\setcounter{tocdepth}{2}
\newpage
\setcounter{footnote}{0}
\linespread{1.1}
\parskip 4pt

{}~
{}~

\section{Introduction}

The discovery of integrability in the planar limit of ${\cal N}=4$ super Yang-Mills theory\cite{Beisert:2010jr} has provided
important lessons into gauge/gravity duality\cite{Maldacena:1997re,Gubser:1998bc,Witten:1998qj}.
The planar spectrum can be computed exactly to all order in $\lambda$ and it can be matched to string 
theory - a remarkable achievement\cite{Gromov:2009tv}.

By restricting to the planar limit, we are necessarily restricting attention to operators with a dimension that obeys 
$\Delta^2\ll N$\cite{Balasubramanian:2001nh}.
This is a tiny part of the theory and to properly understand gauge/gravity duality we will presumably have to consider
operators with a dimension of order $N$ or even order $N^2$.
These have a sensible physical interpretation as 
branes\cite{Balasubramanian:2001nh,Berenstein:2003ah,Corley:2001zk,Aharony:2002nd} and new 
geometries\cite{Lin:2004nb} respectively.
The study of these large dimension operators is challenging.
In general, we do not expect any integrability.
Further, the usual description of the large $N$ expansion as a genus expansion for sums of ribbon graphs is not
a valid description and all the known lore of large $N$ must be revisited.

In this study we consider correlation functions involving operators with a dimension of order $N_1$ in a supersymmetric 
${\cal N} = 6$ Chern-Simons-matter theory with gauge group $U(N_1)_k\times U(N_2)_{-k}$, where $k$ denotes 
the Chern-Simons level and we assume that $N_1\sim N_2$.
Our notation is $N_1\ge N_2$.
There is an AdS$_4$/CFT$_3$ duality which relates this Chern-Simons matter theory to type IIA string theory on 
AdS$_4 \times$CP$^3$ with non-zero background fluxes.
There are $N_2$ units of RR four-form flux through AdS$_4$, $k$ units of RR two-form flux through a CP$^1\subset$CP$^3$
and NS B-field $B_2$ with non-trivial holonomy
\begin{equation}
{1\over 2\pi}\int_{{\rm CP}^1\subset {\rm CP}^3} B_2 = {N_1 - N_2\over k}
\end{equation}
For $N_1=N_2$ the Chern-Simons-matter theory is known as ABJM theory\cite{Aharony:2008ug}.
The general case ($N_1\ne N_2$) is denoted ABJ theory\cite{Aharony:2008gk}. 
The fields can be rescaled by powers of ${1\over k}$ so that all interaction vertices are suppressed by powers of ${1\over k}$.
Thus, the level $k$ plays the role of the coupling constant and large $k$ is weak coupling.
The planar limit is given by
\begin{equation}
k,N\to\infty\qquad \lambda\equiv {N\over k}={\rm fixed}
\end{equation}
Integrability makes an appearance in this limit\cite{abjmqsc}.
The theory has two gauge fields, one in the adjoint of $U(N_1)$ and one in the adjoint of $U(N_2)$, four complex scalars
and four Majorana fermions.
The scalars and fermions are both in the $N_1\times\bar N_2$ or $\bar N_1\times N_2$ of $U(N_1)\times U(N_2)$.
We study determinant, sub determinant and permanent operators constructed using only the four complex scalars.

Our goal is to generalize the recently developed techniques of \cite{Jiang:2019xdz,Chen:2019gsb} in the 
AdS$_5$/CFT$_4$ setting to the AdS$_4$/CFT$_3$ setting.
Denote the four complex scalar fields by $A_i$ and $B_i$, $i=1,2$. 
Let $a=1,...,N_1$ be a gauge group index for $U(N_1)$ and let $\alpha=1,...,N_2$ be a gauge group index for $U(N_2)$.
Indicating gauge indices we have $(A_i)^a_\alpha$ and $(B_i)^a_\alpha$.
In this study we work entirely in the free theory.
The free field theory action is given by
\begin{eqnarray}
S=k\int d^3 x \left( \partial_\mu (A_i)^a_\alpha \partial^\mu (A_i^\dagger)_a^\alpha
+\partial_\mu (B_i)^a_\alpha \partial^\mu (B_i^\dagger)_a^\alpha\right)
\end{eqnarray}
Notice that the composite field $(\phi_{ij})^a_b=(A_i)^a_\alpha (B_j^\dagger)^\alpha_b$ transforms in the adjoint of 
$U(N_1)$.
The heavy operators that we are interested in can be described as Schur polynomials in the matrix 
$Z=\phi_{11}$ or $Z^\dagger$ \cite{Dey:2011ea,Chakrabortty:2011fd,Caputa:2012dg}.
As in ${\cal N}=4$ super Yang-Mills theory, the Schur polynomials provide a complete basis for local operators
constructed from $Z$ and they diagonalize the free field theory two point 
function\cite{Dey:2011ea,Chakrabortty:2011fd,Caputa:2012dg}
\begin{equation}
\langle\chi_R(x_1)\chi_S(x_2)^\dagger\rangle = \delta_{RS}f_R(N_1)f_R(N_2)\left({1\over 4\pi k|x_1-x_2|}\right)^n
\label{SchurAnswer}
\end{equation}
where $R$ is a Young diagram with $n$ boxes, i.e. $R\vdash n$.
$f_R(N)$ is a product of factors, one for each box in $R$, with the factor for a box in row $i$ and column $j$ given by
$N-i+j$.
The stringy exclusion principle is implemented by requiring that $R$ has no more than $N_2$ rows.
We will also consider restricted Schur polynomials in the ABJM theory, constructed using $\phi_{11}$ and 
$\phi_{12}$\cite{deMelloKoch:2012kv}.
Operators constructed using $n_{11}$ $\phi_{11}$ fields and $n_{12}$ $\phi_{12}$ fields are labeled by three Young
diagrams, $r\vdash n_{11}$, $s\vdash n_{12}$ and $R\vdash n$ with $n=n_{11}+n_{12}$.
The pair $(r,s)$ label an irreducible representation that can be obtained from the irreducible representation $R$ of
$S_n$ after restricting to the $S_{n_{11}}\times S_{n_{12}}$ subgroup.
The representation $(r,s)$ may appear more than once after restricting and consequently we need a multiplicity label to
distinguish the different copies.
The relevant two point function is given by
\begin{equation}
\langle\chi_{R,(r,s)\alpha\beta}(x_1)\chi_{T,(t,u)\gamma\tau}^\dagger(x_2)\rangle =
\delta_{RS}\delta_{rt}\delta_{su}\delta_{\alpha\gamma}\delta_{\beta\tau}
{{\rm hooks}_R f_R(N_1)^2\over {\rm hooks}_r{\rm hooks}_s}
\left({1\over 4\pi k|x_1-x_2|}\right)^n
\end{equation}
where ${\rm hooks}_t$ stands for the product of hook lengths in Young diagram $t$ and the indices
$\alpha,\beta,\gamma,\tau$ are multiplicity labels.
For a careful and elegant treatment of the effects of the stringy exclusion principle see \cite{Pasukonis:2013ts}.

The string theory duals to these heavy operators are giant gravitons branes in IIA string theory.
Operators labeled by Young diagrams with $O(1)$ long rows (of length $\sim N_1\sim N_2$) correspond to
dual giant gravitons, given by $D2$ branes wrapping an 
S$^2\subset$AdS$_4$\cite{Nishioka:2008ib,Berenstein:2009sa,SheikhJabbari:2009kr,Hamilton:2009iv}.
Operators labeled by Young diagrams with $O(1)$ long columns (of length $\sim N_1\sim N_2$) correspond to
giant gravitons, given by $D4$ branes wrapping a four manifold in 
CP$^3$\cite{Giovannoni:2011pn,Hirano:2012vz,Lozano:2013sra}.

The paper is organized as follows:
In Section \ref{Determinants} we discuss correlation functions involving determinants. 
We start with a discussion of maximal giant gravitons in the ABJM theory and then generalize the discussion to 
general giant gravitons in both ABJM and ABJ theory.
In Section \ref{Permanents} the discussion is generalized to correlation functions of permanents, relevant for
dual giant gravitons.
This is followed in Section \ref{AddingMoreMatrices} with a discussion of restricted Schur polynomials which are
dual to giant gravitons carrying more than one angular momentum.
Following \cite{Jiang:2019xdz,Chen:2019gsb} we explain in Section \ref{GraphDuality}, that the effective theories 
that we obtain can be understood in terms of a graph duality proposed by \cite{Rajesh}.
In an attempt to gain further insight into the $\rho$ theory, we consider a saddle point evaluation of the $\rho$ integral in 
Section \ref{saddleanalysis}, which allows us to obtain the correct leading contribution to the correlators in the large $N$
limit.
An interesting feature of this analysis, for the ABJ theory, is the existence of a pair of saddle points related by parity.
Finally in section \ref{Discussion} we discuss our results and draw some conclusions.

\section{Correlators involving Determinants and Subdeterminants}\label{Determinants}

The maximal giant gravitons in the Anti-de Sitter spacetime are dual to determinant operators in the CFT so that we will
refer to the determinant operators as maximal giant gravitons. 
We are interested in computing the correlation function of $Q$ maximal giant gravitons, located at positions $x_A$, 
for $A=1,2,...,Q$.
The giant at $x_A$ is given by the Schur polynomial $\chi_{(1^{N_1})}(x_A)$.
The Schur polynomial located at $x_A$ is constructed using the field $({\cal Z}_A)^a_b(x_A)$, which is a linear combination
of products of pairs of the complex scalar fields, each transforming in the adjoint of $U(N_1)$.
The only assumption we make is that
\begin{equation}
\left\langle ({\cal Z}_K)^a_b(x_K)({\cal Z}_K)^c_d(x_K)\right\rangle =0\label{Constr}
\end{equation}
which ensures that our composite operator is free of UV divergences.
For simplicity, to start, consider the ABJM theory.
Introduce two sets of fermionic vectors, $\chi^a,\bar\chi_a$ and
$\psi^\alpha,\bar\psi_\alpha$ and note that 
\begin{eqnarray}
\int [d^{N_1}\bar\chi d^{N_1}\chi][d^{N_1}\bar\psi d^{N_1}\psi] 
\bar\psi_{\beta_1} \psi^{\alpha_1}\cdots\bar\psi_{\beta_{N_1}}\psi^{\alpha_{N_1}}
\bar\chi_{b_1}\chi^{a_1}\cdots\bar\chi_{b_{N_1}}\chi^{a_{N_1}}\cr
=\sum_{\sigma,\rho\in S_{N_1}} \chi_{(1^{N_1})}(\sigma)\chi_{(1^{N_1})}(\rho)
\sigma^{\bf \alpha}_{\bf \beta}\rho^{\bf a}_{\bf b}
\end{eqnarray}
where
\begin{equation}
\sigma^{\bf \alpha}_{\bf \beta}=\delta^{\alpha_1}_{\beta_{\sigma(1)}}\cdots \delta^{\alpha_{N_1}}_{\beta_{\sigma(N_1)}}
\qquad\qquad
\rho^{\bf a}_{\bf b}=\delta^{a_1}_{b_{\rho(1)}}\cdots \delta^{a_{N_1}}_{b_{\rho(N_2)}}
\end{equation}
The label $(1^{N_1})$ stands for a Young diagram with a single column of $N_1$ boxes.
We will now argue that the right hand side of the above identity is the projection operator that appears in the
definition of maximal giant gravitons in the ABJM theory.
Using the Fundamental Orthogonality Relation for matrix elements of irreducible representations\cite{Lederman}, it 
is simple to prove the identity
\begin{equation}
\sum_{\sigma,\rho}\chi_R(\sigma)\chi_S(\rho){\rm Tr}(\sigma A^{\otimes N_1} \rho B^{\dagger\, \otimes N_1})
={\delta_{RS}N_1!\over d_R}\sum_{\psi\in S_{N_1}}\chi_R(\psi)
(AB^\dagger)^{\alpha_{1}}_{\alpha_{\psi(1)}}\cdots (AB^\dagger )^{\alpha_{N_1}}_{\alpha_{\psi(N_1)}}
\end{equation} 
As usual there is a Schur-Weyl duality that (in the most general case of ABJ theory) organizes both the representations of 
$U(N_1)$ and $U(N_2)$.
For the example we are considering here the centralizer is $S_{N_1}$. 
In general we could have $n\ne N_1$ fields and the same identity would hold, after replacing $N_1\to n$.
The centralizer in this more general case is $S_n$ - which swaps $A$s and $B^\dagger$s.
Notice that {\it it is the same $S_n$ that is the centralizer for both $U(N_1)$ and $U(N_2)$ and this is why we get
the $\delta_{RS}$ above.} 
The reader should also note that it is only the symmetric group that played a role in the derivation of the above formula,
so that it is also applicable in the ABJ theory where $N_1\ne N_2$.

Using the identity above we will be able to write an integral representation for the maximal giant graviton correlation functions.
To carry out a general discussion introduce a set of vectors $\cal{Y}_K$ which we dot with $\phi^I = (A_1, A_2, B_1, B_2)$ 
and a set of vectors $\bar{\cal{Y}}_K$ which we dot with $\phi^{I\dag} = (A_1^\dag, A_2^\dag, B_1^\dag, B_2^\dag)$. 
For now keep these vectors general up to the $Q$ conditions $\bar{\cal{Y}}_K \cdot {\cal{Y}}_K = 0$, which ensure  contractions of fields inside the same giant vanish, consistent with (\ref{Constr}).  
The correlation function of $Q$ giant gravitons and a single trace operator ${\cal O}$ can be written as 
\begin{eqnarray}
& & \left\langle \chi_{(1^{N_1})}(x_1)\chi_{(1^{N_1})}(x_2)\cdots \chi_{(1^{N_1})}(x_Q)\,{\cal O}\right\rangle
=\int [d\phi^I d\phi^{I\, \dagger}] \int \prod_{K=1}^Q [d^{N_1}\bar\chi_K d^{N_1}\chi_K][d^{N_1}\bar\psi_K d^{N_1}\psi_K]\cr
&& e^{-k\int d^3 x \left( \partial_\mu  \phi^I \cdot  \partial^\mu (\phi^I)^\dagger
+\frac{1}{k}\sum_{K=1}^Q \delta (x-x_K)(\bar \chi_{aK}( {\cal{Y}_K \cdot \phi^I }   )^a_\alpha\psi_K^\alpha
-\bar \psi_{\alpha K}(\bar{\cal{Y}}_K \cdot (\phi^I)^\dag  )^\alpha_a\chi_K^a )\right)}
{\cal O}(\phi^I,\phi^{I\dagger})  \label{phiAction}\cr
&&
\end{eqnarray}
The first step is to perform the Gaussian integral over the adjoint scalars which leads to
\begin{eqnarray}
&&\left\langle \chi_{(1^{N_1})}(x_1)\chi_{(1^{N_1})}(x_2)\cdots \chi_{(1^{N_1})}(x_Q)\,{\cal O}\right\rangle
=\int \prod_{K=1}^Q \left[d^{N_1}\bar\chi_K d^{N_1}\chi_K \right]\left[d^{N_1}\bar\psi_K d^{N_1}\psi_K \right] 
\cr &&\qquad\qquad e^{\frac{1}{4k\pi} \sum_{K\neq J = 1}^Q \frac{{\bar{\cal{Y}}}_K \cdot {\cal{Y}}_J}{x_{JK}}  \left( \bar \psi_{\alpha K}\psi_J^\alpha  \bar{\chi}_{a J}  \chi_K^a  \right)   }
{\cal O}(S,S^\dagger)
\end{eqnarray}
where
\begin{equation}
(S^I)^a_\alpha = {1\over k}\sum_{K=1}^Q \frac{1}{4 \pi |x-x_K|}\left( {\bar{\cal{Y}}}^I_K \bar \psi_{\alpha K} \chi_K^a\right)
\qquad
(S^{I\dag})^\alpha_a = -{1\over k}\sum_{K=1}^Q \frac{1}{4 \pi |x-x_K|}\left({\cal{Y}}^I_K \bar\chi_{a K} \psi_K^\alpha\right)
\label{MaxGiant_SI}
\end{equation}
We now perform a Hubbard-Stratonovich transformation, introducing a complex matrix $\rho_{JK}$ and replacing the quartic dependence on the fermion vectors with a quadratic dependence.
It is then possible to integrate over the fermionic vectors to obtain
\begin{eqnarray}
&&\left\langle \chi_{(1^{N_1})}(x_1)\chi_{(1^{N_1})}(x_2)\cdots \chi_{(1^{N_1})}(x_Q)\,{\cal O}\right\rangle\cr
&&\qquad\qquad =
\int [d\rho \,d\rho^\dagger]e^{-4\pi{N_1\over\lambda}{\rm Tr}(\rho^\dag\rho )+N_1 {\rm Tr} \log\left( M_1\right)+N_1 {\rm Tr} \log\left(M_2\right)} 
\langle {\cal O}(S,S^\dagger) \rangle_{\chi,\psi}\label{Fnl}
\end{eqnarray}
where 
\begin{equation}
(M_1)_{JK} = -\sqrt{\frac{\bar{\cal Y}_K \cdot {\cal Y}_J }{|x_{KJ}|}} \rho_{KJ} \qquad\qquad
(M_2)_{KJ} = -\sqrt{\frac{\bar{\cal Y}_K \cdot {\cal Y}_J }{|x_{KJ}|}} \rho^\dag_{JK}
\end{equation}
and the measure is normalized so that
\begin{equation}
\int [d\rho\, d\rho^\dagger ] e^{- 4\pi {N_1\over\lambda}{\rm Tr}\left( \rho \rho^\dag \right)} = 1
\end{equation}
The integration over the fermionic vectors contracts the fermions appearing in $S$ and $S^\dagger$.
The result after the contractions are performed is denoted by $\langle {\cal O}(S,S^\dagger) \rangle_{\chi,\psi}$ in (\ref{Fnl}).
These contractions are carried out by applying Wick's theorem as usual with the following contractions
\begin{equation}
\langle \bar{\chi}_{Ka} \chi^b_J  \rangle = \delta^b_a (M_1^{-1})_{KJ} \qquad
\langle \bar{\psi}_{Ka} \psi^b_J  \rangle = \delta^b_a (M_2^{-1})_{KJ}
\end{equation}
Notice that a saddle point evaluation of (\ref{Fnl}) naturally generates the ${1\over N_1}$ expansion.

As a test of (\ref{Fnl}), we will consider the two-point function 
$\langle\chi_{(1^{N_1})}(A B^\dag)\chi_{(1^{N_1})}(A^\dag B)\rangle$
of two giant gravitons in free ABJM theory.
In this case we have
\begin{equation}
\rho = \left( \begin{array}{cc} 0 & z_1 \\ z_2 & 0 \end{array} \right) \qquad
M_1 = -\frac{1}{\sqrt{|x_{12}|}} \left( \begin{array}{cc} 0 & z_2 \\ z_1 & 0   \end{array}\right) \qquad
M_2 = -\frac{1}{\sqrt{|x_{12}|}} \left( \begin{array}{cc} 0 & z_1^* \\ z_2^* & 0   \end{array}\right)
\end{equation}
Using polar coordinates for the complex numbers $z_1,z_2$ we find
\begin{eqnarray}
\langle \chi_{(1^{N_1})}(A B^\dag)(x_1) \,\, \chi_{(1^{N_1})}(A^\dag B)(x_2)\rangle 
&=& \frac{64 N_1^2 \pi^2}{\lambda^2}\frac{1}{|x_{12}|^{2N_1}} \int dr_1 \int dr_2 
e^{-\frac{4\pi N_1}{\lambda}(r_1^2 + r_2^2)} r_1^{2N_1+1} r_2^{2N_1+1}\cr
&=& (N_1!)^2 \left( \frac{1}{4 \pi k|x_{12}| }\right)^{2N_1}
\end{eqnarray}
which agrees with (\ref{SchurAnswer}).  
Now consider the same computation in the free ABJ theory and recall that $N_1 \geq N_2$.  
The only significant difference between the ABJM and ABJ theories is in the initial integral expression for the correlator.  
The relevant integral representation for the ABJ theory is
\begin{eqnarray}
& & \left\langle \chi_{(1^{N_2})}(x_1)\chi_{(1^{N_2})}(x_2)\cdots \chi_{(1^{N_2})}(x_Q)\,{\cal O}\right\rangle
=\int [d\phi^I d\phi^{I\dagger}] \int \prod_{K=1}^Q
[d^{N_1}\bar\chi_K d^{N_1}\chi_K][d^{N_2}\bar\psi_K d^{N_2}\psi_K]\cr
& & e^{-k\int d^3 x \left( \partial_\mu  \phi^I \cdot  \partial^\mu (\phi^I)^\dagger
+\frac{1}{k}\sum_{K=1}^Q \delta (x-x_K)(\bar \chi_{aK}( {\cal{Y}_K \cdot \phi^I }   )^a_\alpha\psi_K^\alpha
-\bar \psi_{\alpha K}(\bar{\cal{Y}}_K \cdot (\phi^I)^\dag  )^\alpha_a\chi_K^a  + \bar{\chi}_{a K} \chi^a_K) \right)}
{\cal O}(\phi^I,\phi^{I\dagger})\cr
&&
\end{eqnarray}
The important difference in the ABJ expression above and the ABJM expression in (\ref{phiAction}) is that there is an extra
term $\bar{\chi}_{a K} \chi^a_K$ appearing in the effective action, needed to ``soak up'' the extra $\chi,\bar\chi$ 
integrations.
It is a simple matter to repeat the analysis above with this new starting point.

The above results for determinants can be generalized to sub determinants.
This corresponds to studying correlators of operators dual to giants gravitons that are not necessarily maximal.
A useful identity is
\begin{eqnarray}
&& \int [d^{N_1}\bar\chi d^{N_1}\chi][d^{N_2}\bar\psi d^{N_2}\psi]
\bar\chi_{b_1}\chi^{a_1}\cdots \bar\chi_{b_n}\chi^{a_n} (\bar\chi\cdot\chi)^{N_1-n}
\bar\psi_{\beta_1}\psi^{\alpha_1}\cdots \bar\psi_{\beta_n}\psi^{\alpha_n} (\bar\psi\cdot\psi)^{N_2-n}\cr
&& = (-1)^{N_1-N_2}(N_1-n)!(N_2-n)! \sum_{\sigma,\rho\in S_n}\chi_{(1^n)}(\sigma)\chi_{(1^n)}(\rho)
\sigma^{\bf \alpha}_{\bf \beta}\rho^{\bf a}_{\bf b}\label{goodId}
\end{eqnarray}
where now
\begin{equation}
\sigma^{\bf \alpha}_{\bf \beta}=\delta^{\alpha_1}_{\beta_{\sigma(1)}}\cdots \delta^{\alpha_n}_{\beta_{\sigma(n)}}
\qquad
\rho^{\bf a}_{\bf b}=\delta^{a_1}_{b_{\rho(1)}}\cdots \delta^{a_n}_{b_{\rho(n)}}
\end{equation}
The right hand side of the identity (\ref{goodId}) is the projection operator needed to define sub determinant operators. 
Thus, the generating function for giant graviton correlators can be written as follows
\begin{eqnarray}
&&\sum_{i_1,i_2,\cdots i_Q=1}^N t_1^{2i_1}\cdots t_Q^{2i_Q}
\left\langle \chi_{(1^{i_1})}(x_1)\cdots \chi_{(1^{i_Q})}(x_Q)\,{\cal O}\right\rangle
=\int [d\phi^I d\phi^{I\dagger}] \int \prod_{K=1}^Q
[d^{N_1}\bar\chi_K d^{N_1}\chi_K][d^{N_2}\bar\psi_K d^{N_2}\psi_K]\cr
&&e^{-k\int d^{3}x\,\left\{ \partial_{\mu}\phi^{I}\partial^{\mu}\phi^{I\dagger}
+{1\over k}\sum_{K=1}^{Q}\delta\left(x-x_{K}\right)\left[\bar{\chi}_{K}\cdot\chi_{K}+\bar{\psi}_{K}\cdot\psi_{K}
+t_{K}\bar{\chi}_{K}\mathcal{Z}_{K}\psi_{K}-t_{K}\bar{\psi}_{K}\bar{\mathcal{Z}}_{K}\chi_{K}\right]\right\}}
{\cal O}(\phi^I,\phi^{I\dagger})\cr
&&
\label{giantGenFunction}
\end{eqnarray}
where we have suppressed the gauge indices and 
\begin{equation}
\mathcal{Z}_{K}\equiv\sum_{I=1}^{4}\mathcal{Y}_{K}^{I}\phi^{I}\qquad\qquad
\bar{\mathcal{Z}}_{K}=\sum_{I=1}^{4}\bar{\mathcal{Y}}_{K}^{I}\phi^{I\dagger}
\end{equation}
After integrating over the $\phi^I$ fields, performing a Hubbard-Stratonovich transformation and then integrating
over the fermionic vectors, we obtain
\begin{multline}
\sum_{i_1,i_2,\cdots i_Q=1}^N t_1^{2i_1}t_2^{2i_2}\cdots t_Q^{2i_Q}
\left\langle \chi_{(1^{i_1})}(x_1)\chi_{(1^{i_2})}(x_2)\cdots \chi_{(1^{i_Q})}(x_Q)\,{\cal O}\right\rangle =\int[d\rho d\rho^{\dagger}] \\  
e^{-4\pi k{\rm Tr}\left(\rho\rho^{\dagger}\right)
+N_1 {\rm Tr}\ln\left[\delta_{JK}-\sqrt{\frac{t_{J}t_{K}\bar{\mathcal{Y}}_{K}\cdot\mathcal{Y}_{J}}{\left|x_{KJ}\right|}}\rho_{KJ}\right]
+N_2 {\rm Tr}\ln \left[ \delta_{KJ}-\sqrt{\frac{t_{J}t_{K}\bar{\mathcal{Y}}_{K}\cdot\mathcal{Y}_{J}}{\left|x_{KJ}\right|}}\rho_{JK}^{\dagger} \right]}\left\langle \mathcal{O}^{I}\left(S^{I}, S^{I\dag}\right)\right\rangle _{\chi,\psi}
\label{eq:Giants_rhoTheory},
\end{multline}
where 
\begin{equation}
\left(S^{I}\right)_{\alpha}^{a}=\frac{1}{4\pi k}\sum_{K=1}^{Q}\frac{t_{K}\bar{\mathcal{Y}}_{K}^{I}\bar{\psi}_{K\alpha}\chi_{K}^{a}}{\left|x-x_{K}\right|},\quad\left(S^{I\dagger}\right)_{a}^{\alpha}=-\frac{1}{4\pi k}\sum_{K=1}^{Q}\frac{t_{K}\mathcal{Y}_{K}^{I}\bar{\chi}_{Ka}\psi_{K}^{\alpha}}{\left|x-x_{K}\right|}.
\label{eq:Giant_SI}
\end{equation}
Here $\left\langle \mathcal{O}^{I}\left(S^{I}, S^{I\dag}\right)\right\rangle _{\chi,\psi}$ is again defined by Wick contracting all pairs 
of $\chi, {\bar \chi}$ and $\psi, {\bar \psi}$ fields according to Wick's theorem, with the basic contraction given by
\begin{equation}
\left\langle \bar{\chi}_{aK}\chi_{L}^{b}\right\rangle =\delta_{a}^{b}\left(M_{1}^{-1}\right)_{KL}\qquad\qquad
\left\langle \bar{\psi}_{\alpha K}\psi_{L}^{\beta}\right\rangle =\delta_{\alpha}^{\beta}\left(M_{2}^{-1}\right)_{KL}
\label{Giant_WickContract}
\end{equation}
with 
\begin{equation}
\left(M_{1}\right)_{JK}=\delta_{JK}-\sqrt{\frac{t_{J}t_{K}\bar{\mathcal{Y}}_{K}\cdot\mathcal{Y}_{J}}{\left|x_{KJ}\right|}}\rho_{KJ}\qquad
\left(M_{2}\right)_{KJ}=\delta_{KJ}-\sqrt{\frac{t_{J}t_{K}\bar{\mathcal{Y}}_{K}\cdot\mathcal{Y}_{J}}{\left|x_{KJ}\right|}}\rho_{JK}^{\dagger}
\end{equation}

As a test of (\ref{eq:Giants_rhoTheory}), we compute the two-point $\left\langle \chi_{(1^{J_{1}})}
\left(AB^{\dagger}\right)\chi_{(1^{J_{2}})}\left(A^{\dagger}B\right)\right\rangle$.
The exact result for this two point function is 
\begin{equation}
\left\langle \chi_{(1^{J_{1}})}\left(AB^{\dagger}\right)(x_1)\chi_{(1^{J_{2}})}\left(A^{\dagger}B\right)(x_2)\right\rangle =\delta_{J_{1}J_{2}}\frac{N_{1}!}{\left(N_{1}-J_{1}\right)!}\frac{N_{2}!}{\left(N_{2}-J_{1}\right)!}
\left( {1\over 4\pi k |x_1-x_2|}\right)^{2 J_1}
\end{equation}

In this case, $\bar{\mathcal{Y}}_{1}\cdot\mathcal{Y}_{1}=\bar{\mathcal{Y}}_{2}\cdot\mathcal{Y}_{2}=0$ and 
$\bar{\mathcal{Y}}_{1}\cdot\mathcal{Y}_{2}=\bar{\mathcal{Y}}_{2}\cdot\mathcal{Y}_{1}=1$. 
Parameterize $\rho$ as 
\begin{equation}
\rho=\begin{pmatrix} 0 & z_{1}\\
z_{2} &0
\end{pmatrix}
\end{equation}
The matrices $M_1$ and $M_2$ are then 
\begin{equation}
M_{1}=\begin{pmatrix}1 & -\sqrt{\frac{t_{1}t_{2}}{\left|x_{12}\right|}}z_{2}\\
-\sqrt{\frac{t_{1}t_{2}}{\left|x_{12}\right|}}z_{1} & 1
\end{pmatrix}\qquad \qquad
M_{2}=\begin{pmatrix}1 & -\sqrt{\frac{t_{1}t_{2}}{\left|x_{12}\right|}} z_{1}^{*}\\
-\sqrt{\frac{t_{1}t_{2}}{\left|x_{12}\right|}} z_{2}^{*} & 1
\end{pmatrix}
\label{Giant_2ptM1M2}
\end{equation}
In this case, using (\ref{eq:Giants_rhoTheory}), the computation boils down to computing the integral
\begin{equation}
\int[dz_{1}dz_1^{\dagger} dz_{2}dz_2^{\dagger}]\,e^{- 4\pi k \left(|z_1|^2+|z_2|^{2}\right)}
\det\left(M_{1}\right)^{N_1}\left(\det M_{2}\right)^{N_2}
\end{equation}
Changing to polar coordinates $z_i=r_ie^{i\theta_i}$ and expanding the integrand as follows
\begin{equation}
\left(\det M_{1}\right)^{N_1}\left(\det M_{2}\right)^{N_2}=
\sum_{l_{1},l_{2}}\binom{N_{1}}{l_{1}}\binom{N_{2}}{l_{2}}
\left(-\frac{t_{1}t_{2}}{4\pi k\left|x_{1}-x_{2}\right|}\right)^{l_{1}+l_{2}}
e^{i\left(l_{1}-l_{2}\right)\left(\theta_{1}+\theta_{2}\right)}
\end{equation}
it is simple to find
\begin{align}
\int[d\rho d\rho^{\dagger}]\,e^{-4\pi k {\rm Tr}\rho\rho^{\dagger}}
\left(\det M_{1}\right)^{N_{1}}\left(\det M_{2}\right)^{N_{2}} 
=\sum_{l}\frac{N_{1}!}{\left(N_{1}-l\right)!}\frac{N_{2}!}{\left(N_{2}-l\right)!}
\left(\frac{t_{1}t_{2}}{4\pi k\left|x_{1}-x_{2}\right|}\right)^{2l}
\label{eq:Gaints_2ptResult}
\end{align}
Extracting the coefficient of the $\left(t_1t_2\right)^{2 J_1}$ term, we reproduce the exact result.

With the effective theory we can compute three point functions.
Consider the correlation function
\begin{equation}
\langle \chi_{(1^{K+J})}(x_1)\chi_{(1^{K})}(x_2) {\cal O}_J(x_3)\rangle
\end{equation}
The Schur polynomial at $x_1$ is constructed using $Z^\dagger$, the Schur at $x_2$ is constructed using $Z$
and ${\cal O}_J$ is ${\rm Tr}(Z^J)$.
Here $K$ is order $N_1\sim N_2$, $J$ is order $1$ and we recall that $Z= A_1B^\dagger_1$.
For generality, work in the ABJ model.
Assume that $K+J$ is smaller than both $N_1$ and $N_2$.
The spacetime  dependence of this correlation function is rather simple
\begin{equation}
\langle \chi_{(1^{K+J})}(x_1)\chi_{(1^{K})}(x_2) {\cal O}_J(x_3)\rangle
={C_{K,J}\over (4\pi k|x_1-x_2|)^{2K}(4\pi k|x_1-x_3|)^{2J}}
\end{equation}
where the coefficient $C_{K,J}$ can be computed in the zero dimensional version of the model, in which case it is given by 
\begin{equation}
C_{K,J}=\langle \chi_{(1^{K+J})}(Z^\dagger)\chi_{(1^{K})}(Z) {\rm Tr}(Z^J)\rangle
\end{equation}
Using the identity
\begin{equation}
{\rm Tr}(Z^J)=\sum_{i=0}^{J-1} (-1)^i\chi_{(J-i,1^i)}(Z)
\end{equation}
we have
\begin{equation}
\chi_{(1^{K})}(Z) {\rm Tr}(Z^J)=(-1)^{J-1}\chi_{(1^{K+J})}(Z)+\cdots
\end{equation}
where $\cdots$ above stands for terms with Schur polynomials that have more than a single column and hence don't
contribute to $C_{K,J}$.
Consequently, we have
\begin{equation}
C_{K,J}=\langle \chi_{(1^{K+J})}(Z^\dagger)\chi_{(1^{K})}(Z) {\rm Tr}(Z^J)\rangle
=(-1)^{J-1}{N_1!\over (N_1-K-J)!}{N_2!\over (N_2-K-J)!} \label{exactCorr}
\end{equation}
so that
\begin{eqnarray}
&&\langle \chi_{(1^{K+J})}(x_1)\chi_{(1^{K})}(x_2) {\cal O}_J(x_3)\rangle
=\cr
&&\qquad\qquad
(-1)^{J-1}{N_1!\over (N_1-K-J)!}{N_2!\over (N_2-K-J)!}{1\over (4\pi k|x_1-x_2|)^{2K}(4\pi k |x_1-x_3|)^{2J}}\cr
&&
\end{eqnarray}
Reproducing this result is a convincing check of the effective theory. We consider a more general case with 
\begin{equation}
\mathcal{O}={\rm Tr}\left(\phi^{I_{1}}\phi^{\dagger\bar{I}_{1}}\phi^{I_{2}}\phi^{\dagger\bar{I}_{2}}\cdots\phi^{I_{J}}\phi^{\dagger\bar{I}_{J}}\right).
\end{equation}
Recall that after integrating over the $\phi^I$ and $\phi^{I \dag}$ fields, they become $S^I$'s and $S^{I \dag}$'s.  
Consequently
\begin{eqnarray}
\mathcal{O}& = & (-1)^J\left(\frac{1}{4\pi k}\right)^{2J}\sum_{K_{1},K_{2}\cdots K_{J}=1}^{Q}
\sum_{\bar{K}_{1},\bar{K}_{2}\cdots\bar{K}_{J}=1}^{Q}t_{K_{1}}t_{\bar{K}_{1}}\cdots t_{K_{J}}
t_{\bar{K}_{J}}\bar{\mathcal{Y}}_{K_{1}}^{I_{1}}\mathcal{Y}_{\bar{K}_{1}}^{\bar{I}_{1}}\cdots
\bar{\mathcal{Y}}_{K_{J}}^{I_{J}}\mathcal{Y}_{\bar{K}_{J}}^{\bar{I}_{J}} \nonumber \\
& & \times\frac{   \bar{\psi}_{K_{1}\alpha_{J}}\chi_{K_{1}}^{a_{1}}\bar{\chi}_{\bar{K}_{1}a_{1}}  
\psi_{\bar{K}_{1}}^{\alpha_{1}}\bar{\psi}_{K_{2}\alpha_{1}}\chi_{K_{2}}^{a_{2}}     
\bar{\chi}_{\bar{K}_{2}a_{2}}\psi_{\bar{K}_{2}}^{\alpha_{2}}\cdots\bar{\psi}_{K_{J}\alpha_{J-1}}
\chi_{K_{J}}^{a_{J}}\bar{\chi}_{\bar{K}_{J}a_{j}}\psi_{\bar{K}_{J}}^{\alpha_{J}}}
{\left|x-x_{K_{1}}\right|\left|x-x_{\bar{K}_{1}}\right|\cdots\left|x-x_{K_{J}}\right|\left|x-x_{\bar{K}_{J}}\right|}  
\end{eqnarray}
Wick contractions of the fermionic fields in ${\cal O}$ are given by (\ref{Giant_WickContract}). At the leading order in $N$ we have
\begin{multline}
\left\langle \mathcal{O}\right\rangle =(-1)^{J-1}\left(\frac{1}{4\pi k}\right)^{2J}N_{1}^{J}N_{2}^{J}
\sum_{K_{1},K_{2}\cdots K_{J}=1}^{Q}\sum_{\bar{K}_{1},\bar{K}_{2}\cdots\bar{K}_{J}=1}^{Q}t_{K_{1}}
t_{\bar{K}_{1}}\cdots t_{K_{J}}t_{\bar{K}_{J}}\bar{\mathcal{Y}}_{K_{1}}^{I_{1}}
\mathcal{Y}_{\bar{K}_{1}}^{\bar{I}_{1}}\cdots\bar{\mathcal{Y}}_{K_{J}}^{I_{J}}
\mathcal{Y}_{\bar{K}_{J}}^{\bar{I}_{J}}\\
\times \frac{\left(M_{1}^{-1}\right)_{K_{1}\bar{K}_{1}}\left(M_{2}^{-1}\right)_{\bar{K}_{1}K_{2}}
\left(M_{1}^{-1}\right)_{K_{2}\bar{K}_{2}}\cdots\left(M_{1}^{-1}\right)_{K_{J}\bar{K}_{J}}
\left(M_{2}^{-1}\right)_{\bar{K}_{J}K_{1}}}{\left|x-x_{K_{1}}\right|
\left|x-x_{\bar{K}_{1}}\right|\cdots\left|x-x_{K_{J}}\right|\left|x-x_{\bar{K}_{J}}\right|}
\end{multline} 
Introducing
\begin{align}
\Phi^{I}\left(x\right) & =\left(\frac{N_{1}}{4\pi k}\right)\mathrm{diag}\left(\frac{t_{1}\bar{\mathcal{Y}}_{1}^{I}}{\left|x-x_{1}\right|},\frac{t_{2}\bar{\mathcal{Y}}_{2}^{I}}{\left|x-x_{2}\right|},\cdots,\frac{t_{Q}\bar{\mathcal{Y}}_{Q}^{I}}{\left|x-x_{Q}\right|}\right)M_{1}^{-1}\\
\bar{\Phi}^{\bar{I}}\left(x\right) & =\left(\frac{N_{2}}{4\pi k}\right)\mathrm{diag}\left(\frac{t_{1}\mathcal{Y}_{1}^{\bar{I}}}{\left|x-x_{1}\right|},\frac{t_{2}\mathcal{Y}_{2}^{\bar{I}}}{\left|x-x_{2}\right|},\cdots,\frac{t_{Q}\mathcal{Y}_{Q}^{\bar{I}}}{\left|x-x_{Q}\right|}\right)M_{2}^{-1}
\end{align}
we can write
\begin{equation}
\left\langle \mathcal{O}\right\rangle =(-1)^{J-1} {\rm Tr}\left(\Phi^{I_{1}}\left(x\right)\bar{\Phi}^{\bar{I}_{1}}\left(x\right)\Phi^{I_{2}}\left(x\right)\bar{\Phi}^{\bar{I}_{2}}\left(x\right)\cdots\Phi^{I_{J}}\left(x\right)\bar{\Phi}^{\bar{I}_{J}}\left(x\right)\right)
\end{equation}
For the correlation function we are considering, we have ${\cal O} = {\rm Tr}\left( (A_1 B_1^\dag)^J \right)$ and 
\begin{equation}
\mathcal{Y}_{1}=\left(0,0,1,0\right)\qquad\bar{\mathcal{Y}}_{1}=\left(1,0,0,0\right)\qquad\mathcal{Y}_{2}=\left(1,0,0,0\right)\qquad\bar{\mathcal{Y}}_{2}=\left(0,0,1,0\right)
\end{equation}
%
%
and hence
\begin{align}
\Phi^{A_{1}} & =\left(\frac{N_{1}}{4\pi k}\right)\begin{pmatrix}\frac{t_{1}}{\left|x_{1}-x_{3}\right|} &0\\
0 & 0
\end{pmatrix}
M_{1}^{-1}=\frac{N_{1}t_{1}}{4\pi k\left|x_{1}-x_{3}\right|}
\begin{pmatrix}\left(M_{1}^{-1}\right)_{11} & \left(M_{1}^{-1}\right)_{12}\\
0 & 0
\end{pmatrix}\\
\bar{\Phi}^{B_{1}} & =\left(\frac{N_{2}}{4\pi k}\right)\begin{pmatrix}\frac{t_{1}}{\left|x_{1}-x_{3}\right|} &0\\
0 & 0
\end{pmatrix}
M_{2}^{-1}=\frac{N_{2}t_{1}}{4\pi k\left|x_{1}-x_{3}\right|}\begin{pmatrix}\left(M_{2}^{-1}\right)_{11} & \left(M_{2}^{-1}\right)_{12}\\
0 & 0
\end{pmatrix}
\end{align}
The expectation value of $\mathcal{O}$ in the large $N_{1}$ and $N_{2}$ limit is 
$\left\langle \mathcal{O}\right\rangle =(-1)^{J-1}{\rm Tr}\left(\left(\Phi^{A_{1}}\bar{\Phi}^{B_{1}}\right)^{J}\right)$.
So we find
\begin{equation}
{\rm Tr}\left(\Phi^{A_{1}}\bar{\Phi}^{B_{1}}\right)^{J}=\left(N_{1}N_{2}\right)^{J}t_{1}^{2J}
\left(\frac{1}{4\pi k\left|x_{1}-x_{3}\right|}\right)^{2J}\left[\left(M_{1}^{-1}\right)_{11}
\left(M_{2}^{-1}\right)_{11}\right]^{J},  \label{Trace}
\end{equation}
where $M_1$ and $M_2$ are given by (\ref{Giant_2ptM1M2}). The correlator is given by multiplying (\ref{Trace}) into the integrand of (\ref{eq:Gaints_2ptResult}) which yields
\begin{eqnarray}
\langle \chi_{(1^{K+J})}(x_1)\chi_{(1^{K})}(x_2) {\cal O}_J(x_3)\rangle=(-1)^{J-1} \sum_l (t_1 t_2)^{2l} t_1^{2J}
\frac{(N_1-J)!}{(N_1 -J-l)!}\cr
\times \frac{(N_2-J) !}{(N_2 - J- l)!} N_1^J N_2^J 
\left( \frac{1}{4 \pi k|x_1 - x_2|}\right)^{2l} \left( \frac{1}{4\pi k|x_1 - x_3|}\right)^{2J} \label{corrAnswer}
\end{eqnarray}
This is very close to the exact answer.  
Recall that we only summed the leading order contribution at large $N$ when integrating over $\chi, \bar{\chi}, \psi$ 
and $\bar{\psi}$.  
The corrections to this answer are of order $\frac{J^2}{N}$.  
To suppress these we must take $N_1 >> J$ and $N_2 >> J$.  In this limit we have
\begin{equation}
(N_i - J)! N_i^J = N_i ! +\cdots
\end{equation}
where $\cdots$ are subleading at large $N_i$ so that (\ref{corrAnswer}) is the correct large $N$ result 
for (\ref{exactCorr}).

\section{Correlators involving Permanents}\label{Permanents}

In the previous section we have considered correlation functions involving determinants, which are dual to giant gravitons.
This section extends the discussion by considering permanents which correspond to dual giant gravitons.
We will develop the discussion for the ABJ theory.
To obtain the corresponding results for ABJM theory, we simply set $N_1=N_2$. 

Introduce two sets of commuting vectors, $\varphi^a,\bar\varphi_a$ and $\xi^\alpha,\bar\xi_\alpha$.  
A useful identity is the following
\begin{equation}
\int [d\bar\varphi d\varphi][d\bar\xi d\xi] e^{-\bar\varphi\cdot\varphi-\bar\xi\cdot\xi}
\varphi^{\alpha_1}\bar\varphi_{\beta_1}\cdots\varphi^{\alpha_n}\bar\varphi_{\beta_n} 
\xi^{a_1}\bar\xi_{b_1}\cdots\xi^{a_n}\bar\xi_{b_n}
=\sum_{\sigma,\rho\,\in\, S_n} \chi_{(n)}(\sigma)\chi_{(n)}(\rho)
\sigma^{\bf \alpha}_{\bf \beta}\rho^{\bf a}_{\bf b}\label{CrrtId}
\end{equation}
where now
\begin{equation}
\sigma^{\bf \alpha}_{\bf \beta}=\delta^{\alpha_1}_{\beta_{\sigma(1)}}\cdots \delta^{\alpha_n}_{\beta_{\sigma(n)}}
\qquad
\rho^{\bf a}_{\bf b}=\delta^{a_1}_{b_{\rho(1)}}\cdots \delta^{a_n}_{b_{\rho(n)}}
\end{equation}
The label $(n)$ denotes a Young diagram that is a single row of $n$ boxes. 
The right hand side of the identity (\ref{CrrtId}) is the projector needed to define permanent operators.
Thus the generating function for the correlators of interest is given by
\begin{eqnarray}
&&\sum_{i_1,i_2,\cdots i_Q=1}^{\infty} t_1^{2i_1}t_2^{2i_2}\cdots t_Q^{2i_Q}
\left\langle \chi_{(i_1)}(x_1)\chi_{(i_2)}(x_2)\cdots \chi_{(i_Q)}(x_Q)\,{\cal O}\right\rangle
=\int [d\phi d\phi^\dagger] \int \prod_{K=1}^Q [d\bar\varphi_K d\varphi_K][d\bar\xi_K d\xi_K]\cr
&&\qquad \qquad e^{-k\int d^{3}x\,\left\{ \partial_{\mu}\phi^{I}\partial^{\mu}\phi^{I\dagger}
+{1\over k}\sum_{K=1}^{Q}\delta\left(x-x_{K}\right)\left[\bar{\varphi}_{K}\cdot\varphi_{K}
+\bar{\xi}_{K}\cdot\xi_{K}-t_{K}\bar{\varphi}_{K}\mathcal{Z}_{K}\xi_{K}
-t_{K}\bar{\xi}_{K}\bar{\mathcal{Z}}_{K}\varphi_{K}\right]\right\}}{\cal O}(\phi^I,\phi^{I\dagger})\cr
&&
\end{eqnarray}
where we have defined
\begin{equation}
\mathcal{Z}_{K}\equiv\sum_{I=1}^{4}\mathcal{Y}_{K}^{I}\phi^{I}\qquad\qquad
\bar{\mathcal{Z}}_{K}=\sum_{I=1}^{4}\bar{\mathcal{Y}}_{K}^{I}\phi^{I\dagger}
\end{equation}
After integrating over the $\phi^I$ fields, performing a Hubbard-Stratonovich transformation and then integrating
over the bosonic vectors, we obtain
\begin{multline}
\sum_{i_1,i_2,\cdots i_Q=1}^{\infty} t_1^{2i_1}t_2^{2i_2}\cdots t_Q^{2i_Q}
\left\langle \chi_{(i_1)}(x_1)\chi_{(i_2)}(x_2)\cdots \chi_{(i_Q)}(x_Q)\,{\cal O}\right\rangle \\  
=\int[d\rho d\rho^{\dagger}]e^{- 4\pi k{\rm Tr}\left(\rho\rho^{\dagger}\right)
-N_1 {\rm Tr}\ln \left[\delta_{JK}+\sqrt{\frac{t_{J}t_{K}\bar{\mathcal{Y}}_{K}\cdot\mathcal{Y}_{J}}{\left|x_{KJ}\right|}}\rho_{KJ}\right]
-N_2 {\rm Tr}\ln \left[ \delta_{KJ}+\sqrt{\frac{t_{J}t_{K}\bar{\mathcal{Y}}_{K}\cdot\mathcal{Y}_{J}}{\left|x_{KJ}\right|}}\rho_{JK}^{\dagger} \right]}\left\langle \mathcal{O}^{I}\left(S^{I}(x)\right)\right\rangle _{\varphi,\xi}
\label{eq:dual_rhoTheory}
\end{multline}
where
\begin{align}
S^{I}{}^a_\alpha  =\frac{1}{4\pi k}\sum_{K=1}^{Q}\frac{t_{K}
\bar{\mathcal{Y}}_{K}^{I}\bar{\xi}_{K\alpha}\varphi_{K}^{a}}{\left|x-x_{K}\right|}\qquad\qquad
S^{I\dagger}{}^\alpha_a =\frac{1}{4\pi k}\sum_{K=1}^{Q}\frac{t_{K}\mathcal{Y}_{K}^{I}
\bar{\varphi}_{Ka}\xi_{K}^{\alpha}}{\left|x-x_{K}\right|}
\end{align}
The integration over the bosonic vectors implies that all $\xi$ and $\varphi$ fields are contracted, indicated in the notation
$\left\langle \mathcal{O}^{I}\left(S^{I}(x)\right)\right\rangle _{\varphi,\xi}$.
These contractions are again evaluated using Wick's theorem with the basic contractions given by
\begin{equation}
\left\langle \bar{\varphi}_{aK}\varphi_{L}^{b}\right\rangle =\delta_{a}^{b}\left(M_{1}^{-1}\right)_{KL}\qquad\left\langle \bar{\xi}_{\alpha K}\xi_{L}^{\beta}\right\rangle =\delta_{\alpha}^{\beta}\left(M_{2}^{-1}\right)_{KL}
\end{equation}
with 
\begin{equation}
\left(M_{1}\right)_{JK}=\delta_{JK}
+\sqrt{\frac{t_{J}t_{K}\bar{\mathcal{Y}}_{K}\cdot\mathcal{Y}_{J}}{\left|x_{KJ}\right|}}\rho_{KJ}
\qquad\qquad
\left(M_{2}\right)_{KJ}=\delta_{KJ}+
\sqrt{\frac{t_{J}t_{K}\bar{\mathcal{Y}}_{K}\cdot\mathcal{Y}_{J}}{\left|x_{KJ}\right|}}\rho_{JK}^{\dagger}
\end{equation}

To test (\ref{eq:dual_rhoTheory}) it is instructive to compute the two-point function of dual giant gravitons.
Using (\ref{SchurAnswer}), we know that (the operator $\chi_{(J_1)}$ is at $x_1$ and $\chi_{(J_2)}$ is at $x_2$)
\begin{equation}
\left\langle \chi_{(J_{1})}\left(AB^{\dagger}\right)\chi_{(J_{2})}\left(A^{\dagger}B\right)\right\rangle 
= \delta_{J_1J_2} \frac{(N_1 + J-1)!}{(N_1 - 1)!} \frac{(N_2 + J-1)!}{(N_2 - 1)!} 
\left(\frac{1}{4\pi k\left|x_{1}-x_{2}\right|}\right)^{2J}
\label{2dg}
\end{equation}
For this example we have $\bar{\mathcal{Y}}_{1}\cdot\mathcal{Y}_{1}=\bar{\mathcal{Y}}_{2}\cdot\mathcal{Y}_{2}=0$ 
and $\bar{\mathcal{Y}}_{1}\cdot\mathcal{Y}_{2}=\bar{\mathcal{Y}}_{2}\cdot\mathcal{Y}_{1}=1$, as well as
\begin{eqnarray}
\rho=\begin{pmatrix} 0 & z_{1}\\
z_{2} &0
\end{pmatrix}\quad
M_{1} =\begin{pmatrix}1 & \sqrt{\frac{t_{1}t_{2}}{\left|x_{12}\right|}}z_{2}\\
\sqrt{\frac{t_{1}t_{2}}{\left|x_{12}\right|}}z_{1} & 1
\end{pmatrix}\quad
M_{2}=\begin{pmatrix}1 & \sqrt{\frac{t_{1}t_{2}}{\left|x_{12}\right|}} z_{1}^{*}\\
\sqrt{\frac{t_{1}t_{2}}{\left|x_{12}\right|}} z_{2}^{*} & 1 \end{pmatrix}
\end{eqnarray}
Using (\ref{eq:dual_rhoTheory}) the computation of the correlator boils down to evaluating the integral
\begin{equation}
\int[dz_{1}dz_1^{\dagger} dz_{2}dz_2^{\dagger}]\,e^{- 4\pi k\left(r_{1}^{2}+r_{2}^{2}\right)}\det\left(M_{1}\right)^{-N_{1}}\left(\det M_{2}\right)^{-N_{2}}
\end{equation}
Moving to polar coordinates for the complex variables $z_1,z_2$ and expanding the integrand
\begin{equation}
\det\left(M_{1}\right)^{-N_1}\left(\det M_{2}\right)^{-N_2}=\sum_{j_{1},j_{2}}\binom{N_{1} + j_1-1}{j_{1}}\binom{N_{2}+j_2-1}{j_{2}} e^{i\left(j_{1}-j_{2}\right)\left(\theta_{1}+\theta_{2}\right)} \left({-r_1 r_2 t_1t_2 \over x_{12}}\right)^{j_1 + j_2}
\end{equation}
we easily find
\begin{eqnarray}
\left\langle \chi_{J_{1}}\left(AB^{\dagger}\right)\chi_{J_{2}}\left(A^{\dagger}B\right)\right\rangle
=\sum_{j} \frac{(N_1 + j -1)!}{(N_1-1)!} \frac{(N_2 + j -1)!}{(N_2-1)!}
\left(\frac{t_1 t_2}{4\pi k\left|x_1-x_2\right|}\right)^{2j} \label{2DualGiants}
\end{eqnarray}
which is the correct result.  
The fact that the powers of $t_1$ and $t_2$ are equal reflects the Kronecker delta $\delta_{J_1J_2}$ in (\ref{2dg}).

Now consider a three-point function involving two dual giant gravitons and a single trace ${\cal O}_J = {\rm Tr}\left( (A^{\dag}B)^J \right)$.
Arguing as we did above (see equation (\ref{exactCorr}) and the argument above it) we find
\begin{eqnarray}
&&\langle \chi_{K+J}(x_1)\chi_{K}(x_2) {\cal O}_J(x_3)\rangle=\cr\cr
&&\qquad {(N_1 + J +K -1)!\over (N_1-1)!}{(N_2 + J + K -1)!\over (N_2-1)!}
{1\over (4\pi k|x_1-x_2|)^{2K}(4\pi k|x_1-x_3|)^{2J}}\cr
&&\label{exactCorrT}
\end{eqnarray}
We want to derive the leading behavior at large $N_1,N_2$ of this expression using our effective theory.
To evaluate $\left\langle \mathcal{O}^{I}\left(S^{I}(x)\right)\right\rangle _{\varphi,\xi}$ appearing in
(\ref{eq:dual_rhoTheory}), we need to Wick contract the $\varphi$ and $\xi$ fields using Wick's theorem
with the basic contractions given by
\begin{eqnarray}
\langle \bar{\xi}_{Ka} \xi^b_J \rangle = \delta^b_a \left(M_1^{-1}\right)_{KJ}   \qquad
\langle \bar{\varphi}_{Ka} \varphi^b_J \rangle = \delta^b_a \left(M_2^{-1}\right)_{KJ}  
\end{eqnarray}
Since the bosonic fields commute, we do not need to track any signs.
We only sum the contractions responsible for the leading large $N$ contribution.
A straightforward computation gives
\begin{equation}
{\rm Tr}\left(\Phi^{A_{1}}\bar{\Phi}^{B_{1}}\right)^{J}
=\left(N_{1}N_{2}\right)^{J}t_{1}^{2J}\left(
\frac{g^{2}}{4\pi\left|x_{1}-x_{3}\right|}\right)^{2J}
\left[\left(M_{1}^{-1}\right)_{11}\left(M_{2}^{-1}\right)_{11}\right]^{J}  \label{Trace2}
\end{equation}
To evaluate the correlator of interest, we need to multiply the above result by the integrand relevant for two giant gravitons
and perform the integral over $\rho,\rho^\dagger$.
The integral is performed exactly as in (\ref{2DualGiants}), the only difference being the replacement $N_i \rightarrow N_i + J$.
The result is in agreement with (\ref{exactCorrT}) for large $N$.

\section{Adding More Matrices}\label{AddingMoreMatrices}

In this section we want to consider heavy operators constructed using two matrices, $\phi_{11}=A_1B_1^\dagger$ 
and $\phi_{12}=A_1B_2^\dagger$.
These heavy operators are restricted Schur polynomials\cite{deMelloKoch:2012kv}.
Constructing operators using more than a single matrix corresponding to giving the giant and dual giant gravitons additional
angular momentum.
The restricted Schur polynomial of interest is ($n=n_{11}+n_{12}$)
\begin{equation}
\chi_{(1^{J_1+J_2}),((1^{J_1}),(1^{J_2}))}(\phi_{11},\phi_{12})
={1\over n_{11}^! n_{12}!}\sum_{\sigma\in S_n}
{\rm Tr}_{((1^{J_1}),(1^{J_2}))}\left(\Gamma_{(1^{J_1+J_2)}}(\sigma)\right)
{\rm Tr}(\sigma\phi_{11}^{\otimes n_{11}} \phi_{12}^{\otimes n_{12}})
\end{equation}
The representation labeled by the Young diagram $(1^{J_1+J_2})$ is one dimensional, so the restriction needed is trivial 
(this is also why we don't need multiplicity labels in the above equation) so that we can write
\begin{equation}
\chi_{(1^{J_1+J_2}),((1^{J_1}),(1^{J_2}))}(\phi_{11},\phi_{12})
={1\over n_{11}^! n_{12}!}\sum_{\sigma\in S_n}
\chi_{(1^{J_1+J_2})}\left(\sigma\right)
{\rm Tr}(\sigma\phi_{11}^{\otimes n_{11}} \phi_{12}^{\otimes n_{12}})  \label{multiTwoPoint}
\end{equation}
The two point function of this restricted Schur polynomial is
\begin{eqnarray}
&&\langle \chi_{(1^{J_1+J_2}),((1^{J_1}),(1^{J_2}))}(\phi_{11},\phi_{12})
\chi_{(1^{J_1+J_2}),((1^{J_1}),(1^{J_2}))}^\dagger(\phi_{11},\phi_{12})\rangle\cr
&&\qquad\qquad\qquad=\left({N!\over (N-J_1-J_2)!}\right)^2 {\left(J_1+J_2\right)!\over J_1!J_2!}
\left({1\over 4\pi k|x_1-x_2|}\right)^{J_1+J_2}
\end{eqnarray}
where $\chi_{(1^{J_1+J_2}),((1^{J_1}),(1^{J_2}))}(\phi_{11},\phi_{12})$ is located at $x_1$ and
$\chi_{(1^{J_1+J_2}),((1^{J_1}),(1^{J_2}))}^\dagger(\phi_{11},\phi_{12})$ is located at $x_2$.
Simple character manipulations\cite{Lederman} lead to the following identity
\begin{multline}
\chi_{(1^{J_1+J_2}),((1^{J_1}),(1^{J_2}))}(\phi_{11},\phi_{12}) \\
={1\over n! n_{11}^! n_{12}!}\sum_{\sigma\in S_n}\chi_{(1^{J_1+J_2})}\left(\sigma\right)\chi_{(1^{J_1+J_2})}\left(\rho\right)
{\rm Tr}(\sigma \, A_1^{\otimes n}\, \rho \, 
(B_1^\dagger)^{\otimes n_{11}}(B_2^\dagger)^{\otimes n_{12}})\label{ChrIdnty}
\end{multline}
This last line implies the following expression for the maximal giants (we assume that $Q$ is even and that the giants
for $K=1,...,Q/2$ are built from $A_1B_1^\dagger$ and $A_1B_2^\dagger$, and the remaining giants are built from
$B_1A_1^\dagger$ and $B_2A_1^\dagger$)
\begin{eqnarray}
&&\sum_{n_1,\cdots,n_Q=0}^N t_1^{n_1}\cdots t_Q^{n_Q}
\langle\chi_{(1^N),((1^{n_1}),(1^{N-n_1}))}(x_1)\cdots \chi_{(1^N),((1^{n_Q}),(1^{N-n_Q}))}(x_Q){\cal O}\rangle\cr
&=&\int [dA_i][dA_i^\dagger][dB_i][dB_i^\dagger] \int \prod_{K=1}^Q
[d^{N_1}\bar\chi_K d^{N_1}\chi_K][d^{N_2}\bar\psi_K d^{N_2}\psi_K]e^{S_{\rm eff}}{\cal O}(Z,Z^\dagger)
\end{eqnarray}
with
\begin{eqnarray}
S_{\rm eff}&=&-k\int d^3 x \left( \partial_\mu A_i\partial^\mu A_i^\dagger
+\partial_\mu B_i \partial^\mu B_i^\dagger\right)\cr
&-&\sum_{K=1}^{Q\over 2}(\bar \chi_{aK}(A_1(x_K))^a_\alpha\psi_K^\alpha
-\bar \psi_{\alpha K}(t_K(B_1^\dagger)_K+(B_2^\dagger)_K)^\alpha_a\chi_K^a)\cr
&-&\sum_{K=1+{Q\over 2}}^{Q}(\bar \chi_{aK}(t_K(B_1)_K+(B_2)_K)^a_\alpha\psi_K^\alpha
-\bar \psi_{\alpha K}(A_1^\dagger)^\alpha_a\chi_K^a)
\end{eqnarray}
For the non-maximal giants we again consider a correlator with the first $\frac{Q}{2}$ giants built from $A_1B_1^\dagger$ and $A_1B_2^\dagger$ while the remaining giants are built from $B_1A_1^\dagger$ and $B_2A_1^\dagger$.  For this we will consider
\begin{eqnarray}
&&\sum_{n_1,\cdots,n_Q=0}^N (t_{1Y})^{i_1} (t_{1Z})^{j_1} \cdots  (t_{QY})^{i_Q} (t_{QZ})^{j_Q} \langle\chi_{(1^{i_1 + j_1}),((1^{i_1}),(1^{j_1}))}(x_1)\cdots \chi_{(1^{i_Q + j_Q}),((1^{i_Q}),(1^{j_Q}))}(x_Q){\cal O}\rangle\cr  \nonumber \\
&=&\int [d\phi^I d\phi^{I\dagger}] \int \prod_{K=1}^Q
[d^{N_1}\bar\chi_K d^{N_1}\chi_K][d^{N_2}\bar\psi_K d^{N_2}\psi_K]\ e^{S_{\rm eff}}
{\cal O}(\phi^I,\phi^{I\dagger})
\end{eqnarray}
where
\begin{eqnarray}
S_{\rm eff}&=&-k\int d^3 x \left( \partial_\mu A_i\partial^\mu A_i^\dagger
+\partial_\mu B_i \partial^\mu B_i^\dagger\right) -  \sum_{K=1}^Q \left( \bar{\chi}_K^a \chi_{a K} + \bar{\psi}_K^{\alpha} \psi_{\alpha K}\right) \nonumber \\
& & 
-\sum_{K=1}^{Q\over 2}(\bar \chi_{aK}(A_1(x_K))^a_\alpha\psi_K^\alpha
-\bar \psi_{\alpha K}(t_{KY}(B_1^\dagger)_K+t_{KZ}(B_2^\dagger)_K)^\alpha_a\chi_K^a)\cr
&-&\sum_{K=1+{Q\over 2}}^{Q}(\bar \chi_{aK}(t_{KY}(B_1)_K+t_{KZ}(B_2)_K)^a_\alpha\psi_K^\alpha
-\bar \psi_{\alpha K}(A_1^\dagger)^\alpha_a\chi_K^a)
\end{eqnarray}
As before we complete the square and integrate over the adjoint scalars to find
\begin{eqnarray}
& & \sum_{n_1,\cdots,n_Q=0}^N (t_{1Y})^{i_1} (t_{1Z})^{j_1} \cdots  (t_{QY})^{i_Q} (t_{QZ})^{j_Q} \langle\chi_{(1^{i_1 + j_1}),((1^{i_1}),(1^{j_1}))}(x_1)\cdots \chi_{(1^{i_Q + j_Q}),((1^{i_Q}),(1^{j_Q}))}(x_K){\cal O}\rangle \nonumber \\
& = &
\int \prod_{K=1}^Q [d^{N_1}\bar\chi_K d^{N_1}\chi_K][d^{N_2}\bar\psi_K d^{N_2}\psi_K] e^{\frac{1}{4 \pi k} \sum_{K\neq J=1}^Q \frac{ (t_{KY} t_{JY} + t_{KZ} t_{JZ})\delta_{(K\leq Q/2 <J)} + \delta_{(J\leq Q/2 <K)}  }{\left|x_{K}-x_{J}\right|}\bar{\psi}_{K\alpha}\psi_{J}^{\alpha}\bar{\chi}_{Ja}\chi_{K}^{a}} \nonumber \\
& & \ \ \ \ \ \ \ \ \ \ \ \ \ \ \ \ \ \ \ \ \ \ \ \ \ \ \ \ \ \ \  \times e^{ - \sum_{K=1}^{Q}\left(\bar{\chi}_{K}\cdot\chi_{K}+\bar{\psi}_{K}\cdot\psi_{K}\right)}
{\cal O}(S^I,S^{I\dagger}),
\end{eqnarray}
where $S^I$ and $S^{I\dag}$ can be still written as (\ref{MaxGiant_SI}) if we define
\begin{align}
\mathcal{Y}_{K}^{I} & =\left\{ \delta_{(K\leq Q/2)},0,t_{KY}\delta_{(K>Q/2)},t_{KZ}\delta_{(K>Q/2)}\right\}, \\
\bar{\mathcal{Y}}_{K}^{I} & =\left\{ \delta_{(K>Q/2)},0,t_{KY}\delta_{(K\leq Q/2)},t_{KZ}\delta_{(K\leq Q/2)}\right\}.
\end{align}
By performing the HS transformation we find
\begin{eqnarray} 
& & \sum_{n_1,\cdots,n_Q=0}^N (t_{1Y})^{i_1} (t_{1Z})^{j_1} \cdots  (t_{QY})^{i_Q} (t_{QZ})^{j_Q} \langle\chi_{(1^{i_1 + j_1}),((1^{i_1}),(1^{j_1}))}(x_1)\cdots \chi_{(1^{i_Q + j_Q}),((1^{i_Q}),(1^{j_Q}))}(x_K){\cal O}\rangle \nonumber \\
& = & \int[d\rho d\rho^{\dagger}]e^{-4\pi k {\rm Tr}\left(\rho\rho^{\dagger}\right)+N {\rm Tr}\ln \left[M_1\right]+N {\rm Tr}\ln \left[ M_2 \right]}\left\langle \mathcal{O}^{I}\left(S^{I}(x), S^{I\dagger}(x)\right)\right\rangle _{\chi,\psi}
\end{eqnarray}
where 
\begin{eqnarray}
(M_1)_{KJ} & = & \delta_{JK}- \sqrt{\frac{ \left(t_{KY} t_{JY} + t_{KZ} t_{JZ}\right)\delta_{(K\leq Q/2 <J)} + \delta_{(J\leq Q/2 <K)} }{\left|x_{KJ}\right|}}\rho_{KJ} \nonumber \\
(M_2)_{JK} & = & \delta_{KJ}-\sqrt{\frac{\left(t_{KY} t_{JY} + t_{KZ} t_{JZ}\right)\delta_{(K\leq Q/2 <J)} + \delta_{(J\leq Q/2 <K)} }{\left|x_{KJ}\right|}}\rho_{JK}^{\dagger}
\end{eqnarray}
As a test of the above result, we will reproduce the two-point function (\ref{multiTwoPoint}).  
This amounts to evaluating the integral
\begin{eqnarray}
& & \int[d\rho d\rho^{\dagger}]e^{-4\pi k\left( z^{*}_1 z_1 + z^{*}_2 z_2 \right)}\left(1 -  \sqrt{ t_{1Y} t_{2Y} + t_{1 Z} t_{2 Z}} \frac{z_1 z_2}{|x_{12}|} \right)^{N} \left(1 - \sqrt{ t_{1Y} t_{2Y} + t_{1 Z} t_{2 Z}} \frac{z_1^{*} z^{*}_2}{|x_{12}|} \right)^{N} \nonumber \\
&=& \int [d\rho d\rho^{\dagger}] e^{-4\pi k(r_1^2 + r_2^2)} \sum_{k_1,k_2=0}^N \left( \begin{array}{c} N \\ k_1 \end{array}\right) \left( \begin{array}{c} N \\ k_2 \end{array}\right) \left(-\frac{\sqrt{  t_{1Y} t_{2Y} + t_{1 Z} t_{2 Z} }}{|x_{12}|} r_1 r_2\right)^{k_1 + k_2} e^{i(k_1 - k_2)(\theta_1 + \theta_2)} \nonumber \\
& = & \sum_{J=0}^N \left( \begin{array}{c} N \\ J \end{array} \right)^2\int [d\rho d\rho^{\dagger}]  e^{-4\pi k(r_1^2 + r_2^2)}  r_1^{2J} r_{2}^{2J} \left( \frac{  t_{1Y} t_{2Y} + t_{1 Z} t_{2 Z} }{x_{12}^2} \right)^J \nonumber \\
& = & \sum_{J=0}^N \left( \frac{N!}{(N-J)!} \right)^2 \frac{J!}{J_1! J_2!} \left( \frac{1}{4 \pi k |x_{12}|}  \right)^{2J} \left( t_{1Y} t_{2Y}\right)^{J_1}\left(t_{1 Z} t_{2 Z} \right)^{J_2} 
\end{eqnarray}
where $J = J_1 + J_2$.

By replacing the fermionic vectors $\psi$ and $\chi$ that appear in the above analysis, one could easily consider operators
dual to dual giant gravitons.
It is also possible to consider restricted Schur polynomials constructed using more than two matrices.
We will not pursue either of these extensions here.

\section{Graph Duality}\label{GraphDuality}

In the case of ${\cal N}=4$ super Yang-Mills theory, the description employing the $\rho$ field was related to the original
description by means of a graph duality\cite{Jiang:2019xdz,Chen:2019gsb} first explored in \cite{Rajesh}.
In this section we will again argue for this conclusion.
The $\rho$-description of giant gravitons in ABJ(M) is described using the following ``effective action''
\begin{eqnarray}
S_{\rm eff}&=& 
4\pi k{\rm Tr}\left(\rho\rho^{\dagger}\right)-N_1 {\rm Tr} \log \left(M_1 \right)-N_2 {\rm Tr}  \log \left( M_2 \right)
\end{eqnarray}
where
\begin{eqnarray}
M_1 = \delta_{JK}-\sqrt{\frac{t_{J}t_{K}\bar{\mathcal{Y}}_{K}\cdot\mathcal{Y}_{J}}{\left|x_{KJ}\right|}}\rho_{KJ} \qquad
M_2 = \delta_{KJ}-\sqrt{\frac{t_{J}t_{K}\bar{\mathcal{Y}}_{K}\cdot\mathcal{Y}_{J}}{\left|x_{KJ}\right|}}\rho_{JK}^{\dagger}
\end{eqnarray}
Consider $Q=2$ for simplicity and again consider Schur polynomials constructed using $Z \equiv A_1B_1^{\dag}$ and 
$Z^{\dag}\equiv B_1 A_1^{\dag}$, in which case we have
\begin{eqnarray}
\rho & = & \left( \begin{array}{cc} 0 & z_1 \\ z_2 & 0 \end{array} \right) \\
M_1 & = & \left( \begin{array}{cc} 1 & -\sqrt{\frac{t_1 t_2 }{|x_{12}|}} z_2 \\ -\sqrt{\frac{t_1 t_2 }{|x_{12}|}} z_1 & 1   \end{array}\right) \qquad
M_2 = \left( \begin{array}{cc} 1 & -\sqrt{\frac{t_1 t_2 }{|x_{12}|}} z_1^{*} \\ -\sqrt{\frac{t_1 t_2 }{|x_{12}|}} z_2^{*} & 1   \end{array}\right)
\end{eqnarray}
The effective action can be expanded as follows
\begin{eqnarray}
S_{\rm eff} &=& 4\pi k \left(|z_1|^2 + |z_2 |^2 \right) - N_1 \log\left(1- \frac{t_1 t_2 }{|x_{12}|} z_1 z_2 \right) - N_2 \log\left(1- \frac{t_1 t_2 }{|x_{12}|} z_1^{*} z_2^{*} \right) \cr
&=& 4\pi k  \left(|z_1|^2 + |z_2 |^2 \right) 
+N_1\sum_{n=1}^\infty{1\over n}\left(\frac{t_1 t_2 }{|x_{12}|} z_1 z_2\right)^n  
+ N_2\sum_{n=1}^\infty{1\over n}\left(\frac{t_1 t_2 }{|x_{12}|} z_1^{*} z_2^{*}\right)^n
\label{rhoAction}
\end{eqnarray}
Our goal is to use this effective action to reproduce the following two-point correlator between giant gravitons
\begin{equation}
\langle \chi_{1^{J}}\left(AB^{\dag}\right) (x_1 )  \chi_{1^{J}}\left( A^{\dag}B \right)(x_2 )  \rangle = \frac{N_1 !}{(N_1 -J)!} \frac{N_2 !}{(N_2 -J)!} \left(\frac{1}{4\pi k |x_{12}|} \right)^{2 J}
\end{equation}
This is an exact result in the free field theory.
It is worth explaining a few rules for how diagrams in the original ABJM/ABJ description map into diagrams of the $\rho$ theory.
First, propagators in the original description are propagators in the $\rho$ description.
Thus, the power of ${1\over k}$ tells us how many $\rho$ propagators there are.
The $\rho$ propagators are oriented so we need an arrow on each propagator.
Second, faces map into vertices. Thus, the power of $N_1$ tells us how many vertices there are that come from the second
term in (\ref{rhoAction}) and the power of $N_2$ tells us how many vertices come from the third term.
All the lines on an $N_1$ vertex point outwards and all lines on an $N_2$ vertex point inwards.
The duality maps each of the original diagrams into a new diagram, i.e. it works diagram by diagram.
Further, it maps connected diagrams into connected diagrams and it preserves the number of disconnected components.

The $\rho$ theory has an infinite number of different vertices from which graphs may be composed.
To see what vertices correspond to a given ribbon graph, decompose the ribbon graph into a set of color loops.
The ribbon graph is made from ribbons of many colors because there are many types of fields.
We will restrict our discussion to the $Z$ field introduced above.
In this case we need two colors, red for $A_1,A_1^\dag$ and blue for $B_1,B_1^\dag$.
Each loop is a face of the original ribbon graph, so that it maps into a vertex of the $\rho$ graph.
Looking at the colors (red or blue) of the edges we can read off the structure of the vertex.
Orientation of the edges is assigned so that the loops are correctly glued back together to form the original ribbon graph.
An example to illustrate this procedure is shown in Figure \ref{FDiag4} below.

\vfill\eject

\begin{figure}[ht]%
\begin{center}
\includegraphics[width=0.7\columnwidth]{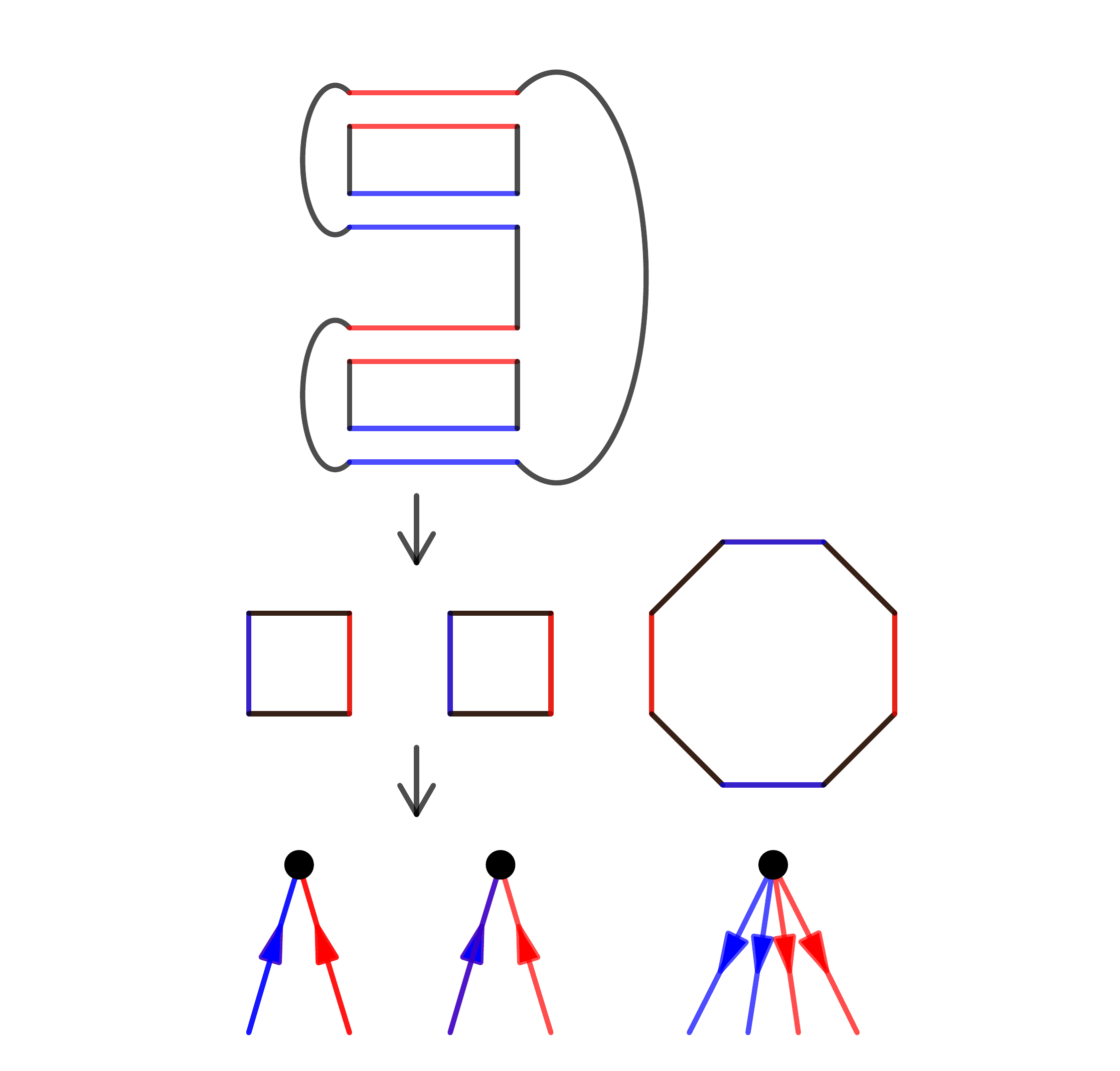}%
\caption{How to read $\rho$ vertices from a ribbon graph.}%
\label{FDiag4}%
\end{center}
\end{figure}

For $J=1$, the two-point correlation function is
\begin{equation}
N_1 N_2 \left(\frac{1}{4 \pi k |x_{12}|} \right)^2 \label{j1result}
\end{equation}
From the powers of $N_i$ we know that the $\rho$ graph has one $N_1$ vertex and one $N_2$ vertex.
From the power of $k^{-1}$ we know that the $\rho$ graph has two propagators.
There is only one diagram with a single $N_1$ vertex, a single $N_2$ vertex and two propagators.
The diagram is shown in the Figure \ref{FDiag1} below.
We use a red line for the $z_1$ propagator and a blue line for the $z_2$ propagator.
there is no non-trivial symmetry factor because the two lines in the graph are inequivalent.
Evaluating this diagram, we  have
\begin{equation}
\frac{1}{4 \pi k } \left(-N_1 \frac{t_1 t_2 }{|x_{12}|} \right)  \times \frac{1}{4 \pi k } \left(-N_2 \frac{t_1 t_2 }{|x_{12}|} \right)
\end{equation}
This is precisely reproduces the exact result (\ref{j1result}).
\begin{figure}[ht]%
\begin{center}
\includegraphics[width=0.4\columnwidth]{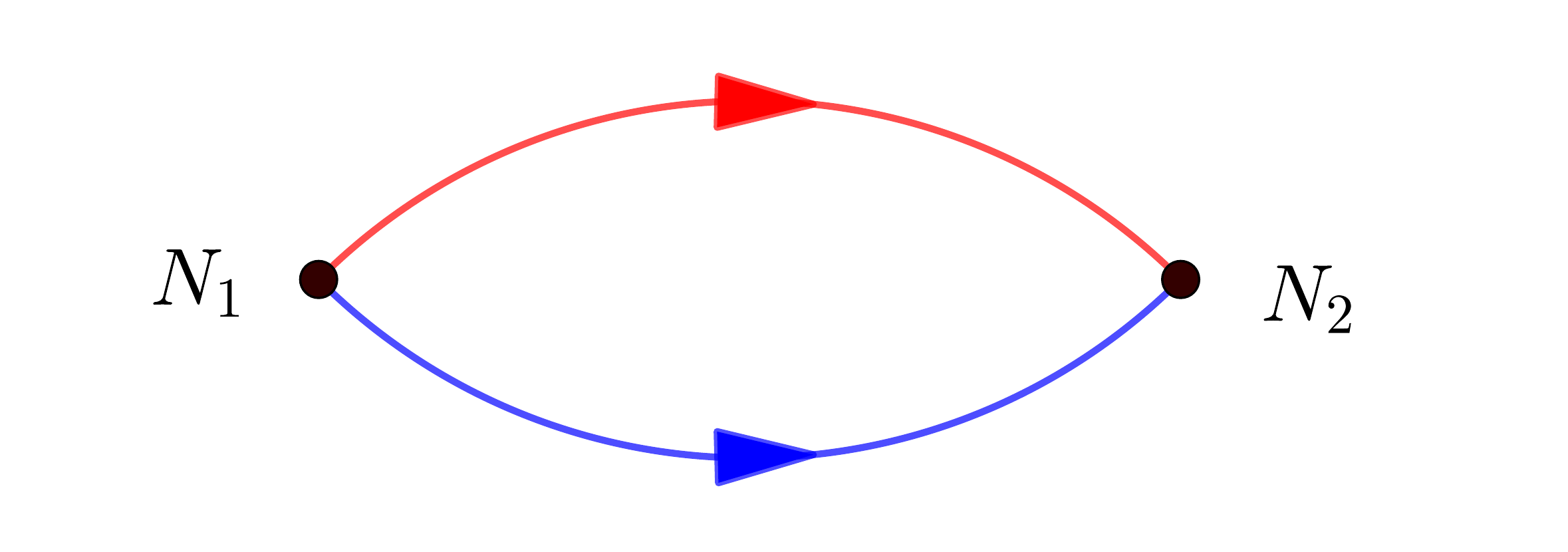}%
\caption{The $\rho$ graph for (\ref{j1result}).}%
\label{FDiag1}%
\end{center}
\end{figure}

For a more nontrivial example,  consider the two-point correlator for $J=2$
\begin{equation}
N_1 (N_1 -1) N_2 (N_2 -1 ) \left(\frac{1}{4 \pi k |x_{12}|} \right)^4  =   \left( N_1^2  N_2^2 - N_1^2 N_2 - N_2^2 N_1 + N_1 N_2 \right) \left(\frac{1}{4 \pi k |x_{12}|} \right)^4    \label{j2giant}
\end{equation}
First, consider the signs.
From the above answer, there are 4 terms which implies that we need to sum 4 $\rho$ graphs. 
Each vertex comes with a $-1$.
The $N_1^2 N_2^2$ graph has 4 vertices and $(-1)^4=1$, the $N_1^2N_2$ and $N_1N_2^2$ graphs each have three 
vertices and $(-1)^3=-1$ and the $N_1N_2$ graph has 2 vertices and $(-1)^2=1$, so the signs are correct.
Each graph has 4 propagators.
The graphs are shown in Figure \ref{FDiag2} below.
\begin{figure}[h]%
\begin{center}
\includegraphics[width=0.9\columnwidth]{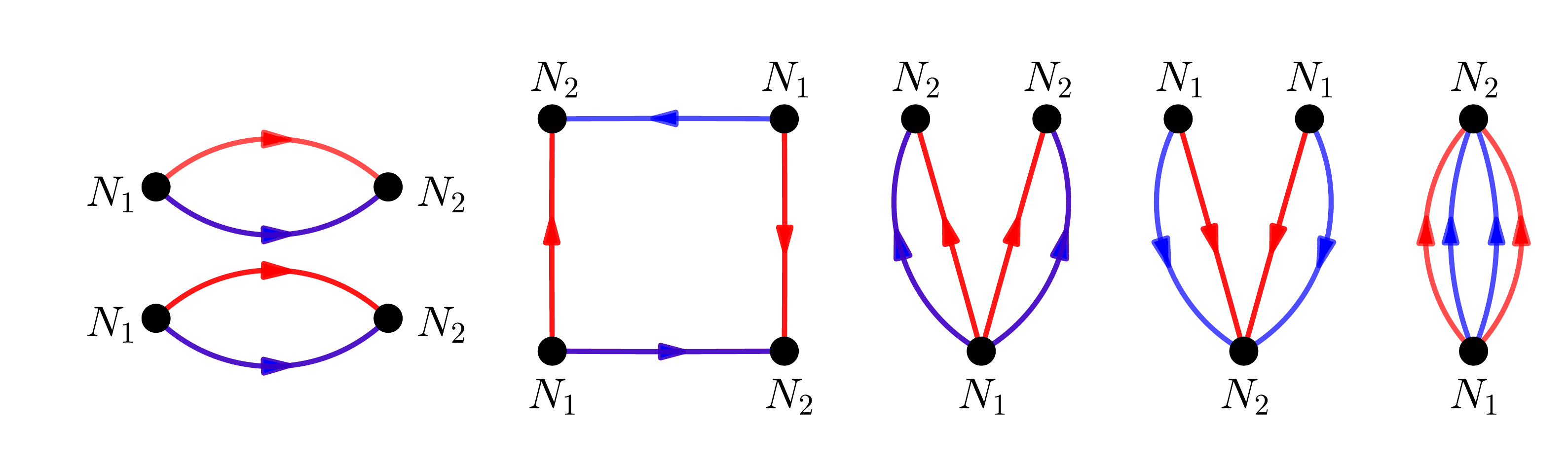}%
\caption{$\rho$ graphs for (\ref{j2giant})}%
\label{FDiag2}%
\end{center}
\end{figure}

For this example there are two distinct graphs that contribute to the leading term. 
The fact that the leading term is a sum of two possible diagrams follows because the leading term
comes from the connected (planar) contribution to $\langle {\rm Tr}(Z^2){\rm Tr}(Z^{\dag 2})$
as well as from the disconnected contribution to $\langle {\rm Tr}(Z)^2{\rm Tr}(Z^{\dag})^2$.
It follows that we need to sum the two diagrams, one connected and one disconnected, that can be formed using
two $N_1$ vertices, two $N_2$ vertices and four propagators.

Finally, consider the $\rho$-description of dual giant gravitons in ABJ(M) theory, which corresponds to the following 
``effective action''
\begin{eqnarray}
S_{\rm eff}&=&4\pi k {\rm Tr}\left(\rho\rho^{\dagger}\right)+N_1 {\rm Tr} \log \left(M_1\right)+N_2 {\rm Tr} \log \left( M_2 \right) 
\end{eqnarray}
where
\begin{eqnarray}
(M_1)_{JK} = \delta_{JK}+\sqrt{\frac{t_{J}t_{K}\bar{\mathcal{Y}}_{K}\cdot\mathcal{Y}_{J}}{\left|x_{KJ}\right|}}\rho_{KJ}, \qquad
(M_2)_{KJ} = \delta_{KJ}+\sqrt{\frac{t_{J}t_{K}\bar{\mathcal{Y}}_{K}\cdot\mathcal{Y}_{J}}{\left|x_{KJ}\right|}}\rho_{JK}^{\dagger}
\end{eqnarray}
Proceeding as above, we find
\begin{eqnarray}
S_{\rm eff} &=& 4\pi k \left(|z_1|^2 + |z_2 |^2 \right) + N_1 \log\left(1- \frac{t_1 t_2 }{|x_{12}|} z_1 z_2 \right) + N_2 \log\left(1- \frac{t_1 t_2 }{|x_{12}|} z_1^{*} z_2^{*} \right) \cr
&=& 4\pi k \left(|z_1|^2 + |z_2 |^2 \right) 
- N_1 \sum_{n=1}^\infty {1\over n}\left(\frac{t_1 t_2 }{|x_{12}|} z_1 z_2\right)^n  - 
N_2\sum_{n=1}^\infty{1\over n} \left( \frac{t_1 t_2 }{|x_{12}|} z_1^{*} z_2^{*}\right)^n\label{dualaction}
\end{eqnarray}
All interaction vertices are positive so that giants and the dual giants have opposite signs for the interaction vertices.
This makes sense because for the symmetric representation all characters are positive and thus all terms in the correlators 
of dual giants are positive.
The two point correlator is
\begin{equation}
\langle \chi_{J }\left(AB^{\dag}\right) (x_1 )  \chi_{J }\left( A^{\dag}B \right)(x_2 )  \rangle = \frac{(N_1  + J -1)!}{(N_1 -1)!} \frac{(N_2 + J-1 )!}{(N_2 -1)!} \left(\frac{1}{4\pi k |x_{12}|} \right)^{2 J}
\end{equation}
For $J=2$ we find
\begin{equation}
N_1 (N_1 +1) N_2 (N_2 +1 ) \left(\frac{1}{4 \pi k |x_{12}|} \right)^4  =   \left( N_1^2  N_2^2 + N_1^2 N_2 + N_2^2 N_1 + N_1 N_2 \right) \left(\frac{1}{4 \pi k |x_{12}|} \right)^4    \label{j2dual}
\end{equation}
All diagrams come with a positive sign which is reproduced by the fact that all vertices in the $\rho$ theory are positive.

\section{Large $N_1$, $N_2$ Saddle Point}\label{saddleanalysis}

The effective action for the $\rho$ theory comes with a factor of $N_i$.
This is an interesting observation because it implies that the loop expansion of the $\rho$ theory is an expansion in
${1\over N_i}$, which is to be contrasted with the description in terms of the original variables, which has the 't Hooft
coupling as the loop expansion parameter.
Recall that the dual holographic description of the CFT has ${1\over N_i}$ for the loop counting parameter. 
Motivated by this observation, we will determine the large $N_i$ saddle points of the ABJ theory in this section.
The analysis is rather interesting: there are two saddles and they are related by parity.
To illustrate this, it is enough to consider the simplest case of $Q=2$.  
Moving to polar-like coordinates $z_i = \sqrt{R_i} e^{i \theta_i}$ the ``effective action'' (see the first line in (\ref{rhoAction})) 
becomes
\begin{eqnarray}
S_{\rm eff} &=& 4 \pi k \left( R_1 + R_2 \right) - N_1 \log\left(1 - \frac{t_1 t_2}{x_{12}} \sqrt{R_1 R_2} e^{i(\theta_1 + \theta_2)} \right)\cr
&-& N_2 \log\left(1 - \frac{t_1 t_2}{x_{12}} \sqrt{R_1 R_2} e^{-i(\theta_1 + \theta_2)} \right)
\end{eqnarray}
Notice that, when $N_1 \neq N_2$ the ``action'' is not hermitian.  
This will be reflected in the saddle point solutions where we have to analytically continue the angular part.  
Of course, the fact that the action is complex is not really a problem because this is not really an action: it simply defines the
generating function of a class of correlation functions of the theory.
There are two saddle points, given by
\begin{eqnarray}
R_{i,\pm} & = & \frac{1}{8 k \pi}\left( N_1 + N_2 \pm \sqrt{(N_1 - N_2)^2 + 4 \chi_{12}^2}\right) \nonumber \\
\left( e^{i(\theta_1 + \theta_2)} \right)_{\pm} & = & \frac{1}{2 \chi_{12}}\left( N_1 - N_2 \pm \sqrt{(N_1 - N_2)^2 + 4 \chi_{12}^2}  \right) 
\end{eqnarray}
where $\chi_{12} =  \frac{4 \pi k x_{12}}{t_1 t_2}$.  
Note that, unless $N_1=N_2$, the angle $\theta_1+\theta_2$ is imaginary.  
By evaluating the action at the saddle points we must recover the leading large $N$ result for the two point correlation function
of giant gravitons.
To reproduce the leading order at large $N$ we need only evaluate the action at the saddle point.
Issues like the normalization of the measure only contribute at subleading order.  
After some simplification we find 

\begin{eqnarray}
e^{-S_{0 +}} & = &  \left(e^{-N_1 - N_2} N_1^{N_1} N_2^{N_2} \chi_{12}^{-(N_1 + N_2)} \right) 
e^{-\sqrt{(N_1 - N_2)^2 + 4 \chi_{12}^2   }}\cr
&&\qquad\qquad\times 
\left(\frac{\sqrt{(N_1 - N_2)^2 + 4 \chi_{12}^2 } + (N_1 - N_2)}{ \sqrt{(N_1 - N_2)^2 + 4 \chi_{12}^2 } - (N_1 - N_2)   }\right)^{\frac{N_1 - N_2}{2}}  \nonumber \\
e^{-S_{0 -}} & = & \left(e^{-N_1 - N_2} N_1^{N_1} N_2^{N_2} \chi_{12}^{-(N_1 + N_2)} \right) e^{\sqrt{(N_1 - N_2)^2 + 4 \chi_{12}^2   }}\cr
&&\qquad\qquad\times 
\left(\frac{\sqrt{(N_1 - N_2)^2 + 4 \chi_{12}^2 } - (N_1 - N_2)}{ \sqrt{(N_1 - N_2)^2 + 4 \chi_{12}^2 } + (N_1 - N_2)   }\right)^{\frac{N_1 - N_2}{2}}   \nonumber 
\end{eqnarray}
These represent the large $N_i$ generating functions.
To extract a specific correlation function, we must read of the coefficient of a given monomial $t_1^n t_2^m$.
The $t_1 t_2$ dependence is contained in the $\chi_{12}$ variable.
To carry out the series expansion it is useful to define $\chi_{12}=\xi_{12}|N_1 - N_2|$.  
Assuming that $N_1 > N_2$, the series expansions yield
\begin{eqnarray}
e^{-S_{0 -}} & = & e^{-2 N_2} \frac{N_{1}^{N_1} N_{2}^{N_2}}{(N_1 - N_2)^{N_1 - N_2}}\frac{1}{(\chi_{12})^{2 N_2}} \nonumber \\
& & + e^{-2 N_2} N_1^{N_1} N_2^{N_2} (N_1 - N_2)^{-N_1 + N_2 - 1} \frac{1}{\chi_{12}^{2 N_2 - 2}} \nonumber \\
& & + \frac{1}{2!} e^{-2 N_2} \frac{N_{1}^{N_1} N_{2}^{N_2}(N_1 - N_2 - 1)  }{(N_1 - N_2)^{N_1 - (N_2-3)}}\frac{1}{(\chi_{12})^{2 N_2 - 4}} + \cdots \nonumber \\
e^{-S_{0 +}} & = & e^{-2 N_1} N_1^{N_1} N_2^{N_2} (N_1 - N_2)^{N_1 - N_2} \frac{1}{\chi_{12}^{2 N_1}} \nonumber \\
& & - e^{-2 N_1} N_1^{N_1} N_2^{N_2} (N_1 - N_2)^{N_1 - N_2 - 1} \frac{1}{\chi_{12}^{2 N_1 - 2}} \nonumber \\
& & + \frac{1}{2!} e^{-2 N_1} N_1^{N_1} N_2^{N_2} (N_1 - N_2)^{N_1 - N_2 - 3} (1 + N_1 - N_2) \frac{1}{\chi_{12}^{2 N_1 - 4}} + \cdots
\end{eqnarray}
These results reproduce the correct leading $N_1, N_2$ behavior of the giant graviton two point function. 
Indeed, taking the coefficient of the third order term from the first expansion, for example, we have
\begin{eqnarray}
\frac{1}{2!} e^{-2 N_2} \frac{N_{1}^{N_1} N_{2}^{N_2}(N_1 - N_2 - 1)  }{(N_1 - N_2)^{N_1 - (N_2-3)}} 
&\approx& e^{-2 N_2} \frac{N_{1}^{N_1} N_{2}^{N_2}  }{(N_1 - (N_2-2) )^{N_1 - (N_2-2)} 2!}\cr 
&\approx& \frac{N_1 ! N_2 !}{(N_1 - (N_2-2) )! 2! }
\end{eqnarray}
where we used Stirling's approximation in the last approximation.  
Notice that the two answers are related by $e^{-S_{0-}}=\left.(-1)^{N_1-N_2}e^{-S_{0 +}}\right|_{N_1 \leftrightarrow N_2}$.  The first series thus represents the correct large $N_1 \geq N_2$ expansion while the second represents the correct large 
$N_2 \geq N_1$ expansion.  
The swap $N_1\leftrightarrow N_2$ is accomplished by parity.
Thus, the two saddles are related by a parity transformation.  
This is exactly what we expect from the breaking of the discrete $Z_2$ parity symmetry when $N_1 \neq N_2$.  

\vfill\eject

\section{Discussion}\label{Discussion}

The basic result of this article is an efficient approach to the computation of correlation functions involving
operators corresponding to giant gravitons and dual giant gravitons, as well as traces, in the ABJM and ABJ theories.
This generalizes results developed in the setting of ${\cal N}=4$ super Yang-Mills theory\cite{Jiang:2019xdz,Chen:2019gsb}.
The derivation of this effective description makes use of a novel identity following from character orthogonality to obtain 
an integral representation of certain projection operators used to define Schur polynomials.
Then, after integrating over the original matrix variables and performing a Hubbard-Stratonovich transformation, one obtains
a description in terms of a $K\times K$ matrix for a collection of $K$ giant or dual giant gravitons. 
The resulting effective descriptions have ${1\over N}$ as the loop counting parameter.
Since ${1\over N}$ controls quantum corrections in the dual gravitational description, this strongly suggests the effective
description is relevant for understanding the holography of the ABJM and ABJ theories.

There are a number of immediate directions that warrant further study.
Our analysis has been restricted to the free field theory.
It would be interesting to consider loop corrections.
Loop corrections to restricted Schur polynomials in ABJM theory have been considered in \cite{Koch:2014csa}.
In addition, the analysis of \cite{Jiang:2019xdz} has suggested that integrability maybe present for correlation functions
involving two determinants and a single trace operator.
A study of loop corrections may establish a similar result in the ABJM/ABJ theories.

Thanks to the fact that there are many different composite adjoint matrices that can be constructed from the ABJM/ABJ
fields, there are many different restricted Schur polynomials one could consider\cite{deMelloKoch:2012kv}.
It would be interesting to develop effective descriptions for this large class of operators.

{\vskip 0.5cm}

\noindent
\begin{centerline} 
{\bf Acknowledgements}
\end{centerline} 

This work is supported by the South African Research Chairs Initiative of the Department of Science and Technology and 
National Research Foundation of South Africa as well as by funds received from the National Institute for Theoretical 
Physics (NITheP).

\appendix

\section{Fermion Measure Conventions}

In this Appendix we spell out our conventions for the fermion measure.
The conventions of this paper agree with those of our last paper \cite{Chen:2019gsb}.
Namely, we use
\begin{equation}
\int d\bar{\psi}_{N}\cdots d\bar{\psi}_{1}\int d\psi^{1}\cdots d\psi^{N}\bar{\psi}_{1}\psi^{1}\cdots\bar{\psi}_{N}\psi^{N}=\left(-1\right)^{N}
\end{equation}
With this convention the Gaussian integral is given by
\begin{equation}
\int d^{N}\bar{\psi}d^{N}\psi\,e^{-\bar{\psi}M\psi}=\det\left(M\right)
\end{equation}

Another commonly used convention, which differs by a phase at most, is 
\begin{equation}
\int d\psi^{1}d\bar{\psi}_{1}\cdots d\psi^{N}d\bar{\psi}_{N}\,\bar{\psi}_{1}\psi^{1}\cdots\bar{\psi}_{N}\psi^{N}=1
\end{equation}
Had we used this convention, the Gaussian integral would become 
\begin{equation}
\int d^{N}\psi d^{N}\bar{\psi}\,e^{\bar{\psi}M\psi}=\det\left(M\right)
\label{eq:A_FermionGaussIntegral}
\end{equation}
With this convention some equations would differ by a signs. 
For example, the sign factor in (\ref{goodId}) disappears
\begin{multline}
\int[d\chi d\bar{\chi}][d\psi d\bar{\psi}]\bar{\chi}_{b_{1}}\chi^{a}\cdots\bar{\chi}_{b_{n}}\chi^{a_{n}}\left(\bar{\chi}\cdot\chi\right)^{N_{1}-n}\bar{\psi}_{\beta_{1}}\psi^{\alpha_{1}}\cdots\bar{\psi}_{\beta_{n}}\psi^{\alpha_{n}}\left(\sum_{i}\bar{\psi}_{i}\psi^{i}\right)^{N_{2}-n}\\
=\left(N_{1}-n\right)!\left(N_{2}-n\right)!\sum_{\sigma,\rho\in S_{n}}\chi_{(1^{n})}\left(\sigma\right)\chi_{(1^{n})}\left(\rho\right)\sigma_{\mathbf{\beta}}^{\mathbf{\alpha}}\sigma_{\mathbf{b}}^{\mathbf{a}}
\end{multline}
The generating function (\ref{giantGenFunction}) would become
\begin{multline}
t_{1}^{2i_{1}}\cdots t_{Q}^{2i_{Q}}\left\langle \chi_{(1^{i_{1}})}\left(x_{1}\right)\cdots\chi_{(1^{i_{Q}})}\left(x_{Q}\right)\mathcal{O}\right\rangle =\int d\phi^{I}d\phi^{\dagger I}\int\prod_{K}^{Q}[d\chi_{K}d\bar{\chi}_{K}][d\psi_{K}d\bar{\psi}_{K}]\\
e^{-k\int d^{3}x\left[\partial_{\mu}\phi^{I}\partial^{\mu}\phi^{\dagger I}-\frac{1}{k}\sum_{K=1}^{Q}\delta(x-x_{K})\left(\bar{\psi}_{K}\cdot\psi_{K}+\bar{\chi}_{K}\cdot\chi_{K}+t_{K}\bar{\chi}_{K}\mathcal{Z}_{K}\psi_{K}-t_{K}\bar{\psi}\bar{\mathcal{Z}}_{K}\chi\right)\right]}\mathcal{O}\left(\phi^{I},\phi^{I\dagger}\right)
\end{multline}
where $\int\prod_{K=1}^{Q}[d\psi_{K}d\bar{\psi}_{K}]\equiv\int\prod_{K=1}^{Q}\prod_{\alpha=1}^{N}d\psi_{K}^{\alpha}d\bar{\psi}_{K\alpha}$.
Note the sign change before the delta function. 
Because of (\ref{eq:A_FermionGaussIntegral}), the sign difference would disappear in (\ref{eq:Giants_rhoTheory}) 
after the integration over the fermion fields is performed. 
This is expected since our results should be independent of these conventions.

\end{document}